\documentclass[10pt]{revtex4-1}
\usepackage{amssymb,amsmath,amscd}
\usepackage{graphicx,calc,epsfig,pstricks,bbm}
\usepackage{tikz}

\newcommand {\be}{\begin{equation}}
\newcommand {\ee}{\end{equation}}
\newcommand{\ba}{\begin{array}{c}}
\newcommand{\ea}{\end{array}}

\newcommand{\scr}{\scriptscriptstyle}
\newcommand{\vgraph}{\mathfrak{n}}
\newcommand{\cube}{\ba
 \begin{tikzpicture}
\pgfmathsetmacro{\cubex}{0.15}
\pgfmathsetmacro{\cubey}{0.15}
\pgfmathsetmacro{\cubez}{0.15}
\draw (0,0,0) -- ++(-\cubex,0,0) -- ++(0,-\cubey,0) -- ++(\cubex,0,0) -- cycle;
\draw (0,0,0) -- ++(0,0,-\cubez) -- ++(0,-\cubey,0) -- ++(0,0,\cubez) -- cycle;
\draw (0,0,0) -- ++(-\cubex,0,0) -- ++(0,0,-\cubez) -- ++(\cubex,0,0) -- cycle;
\end{tikzpicture}
\ea}

\begin{document}

\title{Quantum Reduced Loop Gravity: a realistic Universe}
\author{Emanuele Alesci}
\email{emanuele.alesci@fuw.edu.pl}
\affiliation{Instytut Fizyki Teoretycznej, Uniwersytet Warszawski, ul. Pasteura 5, 02-093 Warszawa, Poland, EU}
\author{Francesco Cianfrani}
\email{francesco.cianfrani@ift.uni.wroc.pl}
\affiliation{Institute
for Theoretical Physics, University of Wroc\l{}aw, Plac\ Maksa Borna
9, Pl--50-204 Wroc\l{}aw, Poland.}

\begin{abstract}
We describe the quantum flat universe in QRLG in terms of states based at cuboidal graphs with six-valent nodes. We investigate the action of the scalar constraint operator at each node and we construct proper semiclassical states. This allows us to discuss the semiclassical effective dynamics of the quantum universe, which resembles that of LQC. In particular, the regulator is identified with the third root of the inverse number of nodes within each homogeneous patch, while inverse-volume corrections are enhanced. 
\end{abstract}

\maketitle

\section{Introduction}

The quantum description of the universe is one of the main goals of any quantum gravity approach, since it is expected to tame one of the pathologies of General Relativity, the initial cosmological singularity. This is generically realized in minisuperspace, {\it i.e.} after imposing homogeneity (eventually isotropy) such that the spatial metric tensor is described by a fiducial static metric and three (eventually one) time-dependent scale factors. The restriction to minisuperspace simplifies the study of the dynamics and it usually allows us to carry on the quantization procedure further.
 
All these features are realized in Loop Quantum Cosmology (LQC) \cite{Bojowald:2011zzb,Ashtekar:2011ni}. LQC adopts some technical tools of Loop Quantum Gravity (LQG)  \cite{Ashtekar:2004eh,Rovelli:2004tv,Thiemann:2007zz} in order to quantize phase space variables in minisuperspace. Despite the difficulties of LQG to analyze the dynamics of the gravitational field, LQC successes in tracing a singularity-free history for the Universe: the Big Bang scenario is replaced by a Bounce occurring when the energy density of the thermal bath reaches a critical value. Furthermore, the non-trivial behavior of perturbations in LQC provides potentially testable implications in observational cosmology, in particular on the cosmic microwave background radiation spectrum \cite{Bojowald:2011hd,Agullo:2012sh}. 
 
In \cite{Alesci:2012md,Alesci:2013xd} we introduced a new model for cosmology in LQG, Quantum Reduced Loop Gravity (QRLG). This model is based on implementing in the kinematical Hilbert space of LQG the gauge fixing conditions imposing diagonal metric tensor and triads \cite{Alesci:2013xya}. This way we get a reduced Hilbert space, whose elements are based at cuboidal graphs and described at links by $U(1)$ group elements, projected from $SU(2)$ ones. In order to discuss the dynamics, some approximations are performed. In particular, the Hamiltonian after the gauge fixing contains some non local contributions, which essentially depend on spatial gradients of reduced phase-space variables. We neglect these terms, that means we are considering only the ``homogeneous'' part of the whole dynamics. This approximation is well-grounded close to a homogeneous configuration and when the Belinski Lipschitz Kalatnikov conjecture \cite{bkl1,Belinsky:1982pk} holds.  

Hence, we get a quantum description for the Bianchi I model in terms of the basic structures of LQG, {\it i.e.} a graph with attached SU(2) quantum numbers, and the dynamics can be analytically investigated. We started such an investigation with the easiest case, a three-valent node. We found that the semiclassical limit of the scalar constraint acting on such a state reproduces the quantum constraint adopted in LQC \cite{Alesci:2014uha} (see \cite{Brunnemann:2005in,
Brunnemann:2005ip} for earlier attempts to relate LQC with LQG via coherent states techniques). Then in \cite{Alesci:2014rra} (see also\cite{nuovo}) we outlined how a consistent picture is inferred by describing each homogeneous Bianchi I patches in terms of semiclassical states based at a generic graph having $N$ six-valent nodes. In fact, the semiclassical limit of the scalar constraint still reproduces the analogous expression in LQC, but the regulator is now related with the number of nodes inside each patch, which is an observable on a quantum level (see \cite{Gambini:2013ooa}). 
		
In this work, we give in details all the technicalities behind the results presented in \cite{Alesci:2014rra}. At first, we present a brief introduction to LQG in section \ref{LQG} and we outline the role of $SU(2)$ recoupling theory in finding the representation of holonomy operators. Then, in section \ref{III}, we discuss the structure of the kinematical Hilbert space in QRLG. We define reduced phase-space variables after imposing the diagonal gauges for the metric and the triads and we recall how to impose the gauge fixing conditions in the kinematical Hilbert space of LQG, resulting in reduced states based at cuboidal graphs with projected group elements at links. We stress how reduced intertwiners arises and how to construct the operators corresponding to phase space variables. At this point we present the new technical tool of this work, which is the new reduced recoupling rule we adopt to represent the action of holonomy operators. 

In previous papers \cite{Alesci:2012md,Alesci:2013xd,Alesci:2014uha}, the recoupling rule for the tensor product of two basis elements (Wigner matrices) is defined from $SU(2)$ recoupling theory by projecting twice: on the magnetic numbers of the initial group elements and on the spin number of the resulting element. Here, we directly adopt $U(1)$ recoupling theory and we outline how it naturally comes out from the full $SU(2)$ recoupling theory when we restrict to reduced group elements.

In section \ref{IV} we analyze the action of the euclidean scalar constraint operator, which is proportional to the full Hamiltonian in the adopted approximation scheme. We start by considering a state based at a three-valent node and we find that there are only some minor changes with respect to the result presented in \cite{Alesci:2014uha}. However, thanks to the new recoupling rule, the extension to states based at six-valent nodes is straightforward and we explicitly compute the action of the scalar constraint on such states. In section \ref{Scal}, semiclassical states are constructed and the expectation value of the single-node Hamiltonian is evaluated. At first, we present the case of the three-valent node and we point out how we formally get the same result as in \cite{Alesci:2014uha}, but in a more consistent approximation scheme. Then, we extend this analysis to a six-valent node. We get the final result for the semiclassical expectation value of the full Hamiltonian in section \ref{VI}, by summing over all the six-valent nodes of the graph. We peak semiclassical states around a homogeneous configuration and we define collective variables. The final expression of the semiclassical effective Hamiltonian in the large $j$-limit reproduces the analogous in LQC, with the regulator given by the third root of the inverse number of nodes. This is our major result, establishing a fundamental correspondence between the effective semiclassical dynamics in QRLG and LQC (and also in Group Field Theory cosmology \cite{Gielen:2013kla,Gielen:2014uga,Calcagni:2014tga}). We also discuss quantum correction and we show how inverse-volume corrections are enhanced with respect to LQC, thus being a distinctive feature of our model. In section \ref{VII} brief concluding remarks follow.

\section{Loop Quantum Gravity}\label{LQG}

The kinematical Hilbert space of QRLG is constructed out of that of LQG by implementing the gauge-fixing conditions restricting to a diagonal metric tensor and diagonal triads. 
In LQG the kinematical Hilbert space $\mathcal{H}$ is constructed by quantizing gravity phase-space parametrized by holonomies and fluxes and it finally reads 
\be\label{kHs}
\mathcal{H}=\oplus_{\Gamma}^G\mathcal{H}_\Gamma\,,
\ee
$^G\mathcal{H}_\Gamma$ being the gauge-invariant space based at each graph $\Gamma$, {\it i.e.} it is the space of gauge-invariant functions of $L$-copies of the SU(2) group, $L$ being the total number of links within $\Gamma$. The direct sum in \eqref{kHs} extends over all the piece-wise analytic graphs in the spatial manifold. Basis elements in $\mathcal{H}$ are invariant spin networks, whose explicit expression reads
\be
<h |\Gamma,\{j_l\},\{x_n\}>=\prod_{n\in\Gamma} {x_{n}}\cdot   \prod_{l}D^{j_{l}}(h_{l}),
\label{spinnet solite} 
\ee
in which $D^{j_{l}}(h_{l})$ are SU(2) irreducible represetations (Wigner matrices) associated with holonomies $h_l$ along each link $l\in\Gamma$, while ${x_{n}}$ denote the invariant intertwiners implementing SU(2) invariance at each node $n$. The products in \eqref{spinnet solite} extend over the nodes $n\in\Gamma$ and all the links $l$ emanating from each $n$. The symbol $\cdot$ means the proper contraction between the indexes of intertwiners and Wigner matrices.

The action of the relevant operators is defined according with canonical quantization. The fluxes across the dual surfaces to a link $l$ act as the left- or right-invariant vector fields of the SU(2) group based at $l$. As soon as functions of holonomies are concerned, their action is that of multiplicative operators. The representation in the spin network basis is given by expanding first the considered operator in SU(2) irreducible representations and, then, computing the product of each term with the Wigner matrices of the basis elements. This procedure can be performed once a prescription for the product of SU(2) irreps is given; this is SU(2) recoupling theory. The basic law of SU(2) recoupling theory can be represented graphically as follows

\be
\begin{array}{c}
\ifx\JPicScale\undefined\def\JPicScale{1}\fi
\psset{unit=\JPicScale mm}
\psset{linewidth=0.3,dotsep=1,hatchwidth=0.3,hatchsep=1.5,shadowsize=1,dimen=middle}
\psset{dotsize=0.7 2.5,dotscale=1 1,fillcolor=black}
\psset{arrowsize=1 2,arrowlength=1,arrowinset=0.25,tbarsize=0.7 5,bracketlength=0.15,rbracketlength=0.15}
\begin{pspicture}(0,0)(25,20)
\psline(12,9)(12,1)
\psline{-<<}(12,5)(5,5)
\psline{<-}(25,5)(18,5)
\psline(12,9)(18,5)
\psline(18,5)(12,1)
\psline(12,20)(12,12)
\psline{-<<}(12,16)(5,16)
\psline{<-}(25,16)(18,16)
\psline(12,20)(18,16)
\psline(18,16)(12,12)
\rput(22,3){$j_1$}
\rput(22,14){$j_2$}
\end{pspicture}
\end{array}
=
\sum_k 
\begin{array}{c}
\ifx\JPicScale\undefined\def\JPicScale{1}\fi
\psset{unit=\JPicScale mm}
\psset{linewidth=0.3,dotsep=1,hatchwidth=0.3,hatchsep=1.5,shadowsize=1,dimen=middle}
\psset{dotsize=0.7 2.5,dotscale=1 1,fillcolor=black}
\psset{arrowsize=1 2,arrowlength=1,arrowinset=0.25,tbarsize=0.7 5,bracketlength=0.15,rbracketlength=0.15}
\begin{pspicture}(0,0)(26,10)
\psline(12,9)(12,1)
\psline{-<<}(9,5)(2,2)
\psline{<-}(26,8)(21,5)
\psline(12,9)(18,5)
\psline(18,5)(12,1)
\psline{-<<}(9,5)(2,8)
\psline{<-}(26,2)(21,5)
\rput(3,0){$j_1$}
\rput(2,10){$j_2$}
\psline(9,5)(12,5)
\psline(18,5)(21,5)
\rput(24,9){$j_2$}
\rput(24,1){$j_1$}
\rput(10,3){$k$}
\end{pspicture}
\end{array}\label{su2rec}
\ee

where the triangles denote SU(2) group elements, the lines are Kronecker $\delta$'s over magnetic indexes and the three-valent nodes represent Clebsch-Gordan coefficients. The sum extends over all the admissible representations $k$, {\it i.e.} $|j_1-j_2|\leq j\leq j_1+j_2$.

\section{Kinematical Hilbert space in QRLG}\label{III}

In QRLG one chooses the reference frame in which the spatial metric tensor is diagonal along some fiducial directions parametrized by the coordinates $x^i$ $(i=1,2,3)$:
\be
ds^2=\mathcal{N}^2dt^2-dl^2\,\qquad dl^2=a_1^2(dx^1)^2+a_2^2(dx^2)^2+a_3^2(dx^3)^2\,,
\ee
$\mathcal{N}$ being the lapse function while $a_i$ denotes the scale factors, 
and one fixes inverse densitized triads such that they are diagonal too, 
\be
E^i_a=\ell_0^{-2}\, p_i\delta^i_a\,,\qquad |p_1|=\ell_0^2\, a_2a_3\quad |p_2|=\ell_0^2\, a_3a_1\quad |p_3|=\ell_0^2\, a_1a_2\,,
\ee 
$V_0=\ell_0^3$ being the fiducial volume of chosen space region. Other phase space variables are the diagonal components of connections $c_i=\ell_0 A^i_a \delta^a_i$, which in terms of the scale factors read
\be
c_i=\ell_0 \frac{\dot{a}_i}{\mathcal{N}}\,,
\ee
where $\dot{}$ denotes time derivative. 

The gauge-fixing conditions are realized in the kinematical Hilbert space of LQG by restricting the class of admissible graphs to cuboidal ones, having links along the fiducial directions, and by a proper projection of SU(2) group elements. This projection is realized such that at each link $l_i$ along $x^i$ it is attached a representation of the U(1) group obtained by stabilizing the SU(2) group along the internal direction $\vec{u}_l=\vec{u}_i$, where
\be
\vec{u}_1=(1,0,0)\quad \vec{u}_2=(0,1,0)\quad \vec{u}_3=(0,0,1)\,.
\ee
This is implemented by projecting the magnetic indexes of the Wigner matrices inside the expression \eqref{spinnet solite} for SU(2) spin networks on the $SU(2)$ coherent states $|\pm j_{l},  \vec{u}_l \rangle$ having maximum or minimum magnetic number along $\vec{u}_l$. Graphically, we can represent the new basis elements by replacing the Wigner matrices at each link $l$ as follows 

\be
\langle j,n | m,\vec{u}_l\rangle \langle m,\vec{u}_l| D^j(h)| m,\vec{u}_l\rangle \langle m,\vec{u}_l|j,r \rangle
=\begin{array} {c} 
\ifx\JPicScale\undefined\def\JPicScale{1.1}\fi
\psset{unit=\JPicScale mm}
\psset{linewidth=0.3,dotsep=1,hatchwidth=0.3,hatchsep=1.5,shadowsize=1,dimen=middle}
\psset{dotsize=0.7 2.5,dotscale=1 1,fillcolor=black}
\psset{arrowsize=1 2,arrowlength=1,arrowinset=0.25,tbarsize=0.7 5,bracketlength=0.15,rbracketlength=0.15}
\begin{pspicture}(0,0)(68,8)
\psline{|*-}(52,3)(48,3)
\rput(35,8){$\scr{j}$}
\rput{0}(35,3){\psellipse[](0,0)(3,-3)}
\rput(35,3){$h$}
\psline[fillstyle=solid]{-|}(12,3)(16,3)
\pspolygon[](6,6)(12,6)(12,0)(6,0)
\psline[fillstyle=solid](2,3)(6,3)
\rput(9,3){$\scr{R_l}$}
\psline[fillstyle=solid](64,3)(68,3)
\pspolygon[](58,6)(64,6)(64,0)(58,0)
\psline[fillstyle=solid]{|*-}(54,3)(58,3)
\rput(61,3){$\scr{R^{-1}_l}$}
\pspolygon[](42,6)(48,6)(48,0)(42,0)
\rput(45,3){$\scr{R_l}$}
\psline[fillstyle=solid](38,3)(42,3)
\psline{}(32,3)(28,3)
\pspolygon[](22,6)(28,6)(28,0)(22,0)
\rput(25,3){$\scr{R^{-1}_l}$}
\psline[fillstyle=solid]{|-}(18,3)(22,3)
\rput(9,8){$\scr{j}$}
\rput(61,8){$\scr{j}$}
\end{pspicture}
\end{array}\qquad m=\pm j\,,
\label{Dridottamn}
\ee
in which the matrix $R_{l}$ denotes the rotation mapping the direction $\vec{u}_l$ into $\vec{u}_3$ and the breaks means the projections on the maximum or minimum magnetic numbers. Hence, basis elements in QRLG on a given cuboidal graph $\Gamma$ can be written as 
\be
\langle h|\Gamma, {\bf m_l, x_n \bf}\rangle= \prod_{n\in\Gamma}\langle{\bf j_{l}}, {\bf x}_n|{\bf m_{l}},  \vec{{\bf u}}_l \rangle 
\prod_{l} \;{}^l\!D^{j_{l}}_{m_{l} m_{l}}(h_{l}),\quad m_l=\pm j_l
\label{base finale}
\ee
where the products $\prod_{n\in\Gamma}$ and $\prod_{l}$ extend over all the nodes $n\in\Gamma$ and over all the links $l$ emanating from $n$, respectively, while there is no contraction of indexes since all the quantities appearing in \eqref{base finale} are complex numbers. It is worth noting how some nontrivial coefficients are induced at nodes, $\langle{\bf j_{l}}, {\bf x}_n|{\bf m_{l}},  \vec{{\bf u}}_l \rangle$, which behaves as one-dimensional intertwiners in QRLG. The space spanned by \eqref{base finale} is denoted by ${}^{R}\mathcal{H}$.  

Flux operators can be inferred by projecting those of LQG and it turns out that they are nonvanishing only when they are smeared along the associated dual fiducial surfaces, {\it i.e.} ${}^{R}\hat{E}_i(S^j)=0$ for $i\neq j$, and they just read the magnetic index of each projected representation when the the surface and the link are dual: 
\be
{}^{R}\hat{E}_i(S^i){}^l\!D^{j_{l}}_{m_{l} m_{l}}(h_{l})= 8\pi\gamma l_P^2\, m_{l}\,{}^l\!D^{j_{l}}_{m_{l} m_{l}}(h_{l}) \qquad l_i\cap S^i\neq \oslash .\label{redei}
\ee

The investigation on the action of holonomy operators requires to fix a proper recoupling theory. In \cite{Alesci:2014uha} we choose the following recoupling rule 
\be
 \begin {split}
\begin{array}{c}
\ifx\JPicScale\undefined\def\JPicScale{1}\fi
\psset{unit=\JPicScale mm}
\psset{linewidth=0.3,dotsep=1,hatchwidth=0.3,hatchsep=1.5,shadowsize=1,dimen=middle}
\psset{dotsize=0.7 2.5,dotscale=1 1,fillcolor=black}
\psset{arrowsize=1 2,arrowlength=1,arrowinset=0.25,tbarsize=0.7 5,bracketlength=0.15,rbracketlength=0.15}
\begin{pspicture}(0,0)(42,19)
\psline(19,8)(19,0)
\psline{-<<}(19,4)(12,4)
\psline{<-}(32,4)(25,4)
\psline(19,8)(25,4)
\psline(25,4)(19,0)
\psline(19,19)(19,11)
\psline{-<<}(19,15)(12,15)
\psline{<-}(32,15)(25,15)
\psline(19,19)(25,15)
\psline(25,15)(19,11)
\rput(29,2){$j_1$}
\rput(28,13){$j_2$}
\psline{<-|}(42,4)(35,4)
\psline{<-|}(42,15)(35,15)
\psline{|-<<}(9,4)(2,4)
\psline{|-<<}(9,15)(2,15)
\psline(12,16)(12,14)
\psline(32,16)(32,14)
\psline(32,5)(32,3)
\psline(12,5)(12,3)
\rput(6,2){$j_1$}
\rput(6,13){$j_2$}
\rput(38,13){$j_2$}
\rput(38,2){$j_1$}
\end{pspicture}
\end{array}
=
\begin{array}{c}
\ifx\JPicScale\undefined\def\JPicScale{1}\fi
\psset{unit=\JPicScale mm}
\psset{linewidth=0.3,dotsep=1,hatchwidth=0.3,hatchsep=1.5,shadowsize=1,dimen=middle}
\psset{dotsize=0.7 2.5,dotscale=1 1,fillcolor=black}
\psset{arrowsize=1 2,arrowlength=1,arrowinset=0.25,tbarsize=0.7 5,bracketlength=0.15,rbracketlength=0.15}
\begin{pspicture}(0,0)(58,9)
\psline(28,8)(28,0)
\psline(28,8)(34,4)
\psline(34,4)(28,0)
\rput(22,2){$\scr{n_1+n_2}$}
\psline{-<<}(7,4)(0,1)
\psline{-<<}(7,4)(0,7)
\rput(1,-1){$n_1$}
\rput(0,9){$n_2$}
\psline{-|}(7,4)(14,4)
\rput(10,2){$\scr{n_1+n_2}$}
\psline{<-}(58,7)(53,4)
\psline{<-}(58,1)(53,4)
\psline{|-}(46,4)(53,4)
\rput(56,8){$n_2$}
\rput(56,0){$n_1$}
\rput(48,2){$\scr{n_1+n_2}$}
\psline{-<<}(28,4)(18,4)
\psline(18,5)(18,3)
\psline{<-}(42,4)(34,4)
\psline(42,5)(42,3)
\end{pspicture}
\end{array}
\end{split}
\label{oldrec}
 \ee 
in which $n_1=\pm j_1, n_2=\pm j_2$ are the magnetic quantum numbers and a second projection on the reduced Hilbert space forces, in the coupling channel $k$, the magnetic number $n_1+n_2$ to be maximum or minimum, {\it i.e.} $k=|n_1+n_2|$.

In this work, we replace the recoupling theory above with the following one

\be
 \begin {split}
\begin{array}{c}
\ifx\JPicScale\undefined\def\JPicScale{1}\fi
\psset{unit=\JPicScale mm}
\psset{linewidth=0.3,dotsep=1,hatchwidth=0.3,hatchsep=1.5,shadowsize=1,dimen=middle}
\psset{dotsize=0.7 2.5,dotscale=1 1,fillcolor=black}
\psset{arrowsize=1 2,arrowlength=1,arrowinset=0.25,tbarsize=0.7 5,bracketlength=0.15,rbracketlength=0.15}
\begin{pspicture}(0,0)(42,19)
\psline(19,8)(19,0)
\psline{-<<}(19,4)(12,4)
\psline{<-}(32,4)(25,4)
\psline(19,8)(25,4)
\psline(25,4)(19,0)
\psline(19,19)(19,11)
\psline{-<<}(19,15)(12,15)
\psline{<-}(32,15)(25,15)
\psline(19,19)(25,15)
\psline(25,15)(19,11)
\rput(29,2){$j_1$}
\rput(28,13){$j_2$}
\psline{<-|}(42,4)(35,4)
\psline{<-|}(42,15)(35,15)
\psline{|-<<}(9,4)(2,4)
\psline{|-<<}(9,15)(2,15)
\psline(12,16)(12,14)
\psline(32,16)(32,14)
\psline(32,5)(32,3)
\psline(12,5)(12,3)
\rput(6,2){$j_1$}
\rput(6,13){$j_2$}
\rput(38,13){$j_2$}
\rput(38,2){$j_1$}
\end{pspicture}
\end{array}
=
\begin{array}{c}
\ifx\JPicScale\undefined\def\JPicScale{1}\fi
\psset{unit=\JPicScale mm}
\psset{linewidth=0.3,dotsep=1,hatchwidth=0.3,hatchsep=1.5,shadowsize=1,dimen=middle}
\psset{dotsize=0.7 2.5,dotscale=1 1,fillcolor=black}
\psset{arrowsize=1 2,arrowlength=1,arrowinset=0.25,tbarsize=0.7 5,bracketlength=0.15,rbracketlength=0.15}
\begin{pspicture}(0,0)(42,12)
\psline(19,10)(19,2)
\psline{-<<}(19,6)(12,6)
\psline{<-}(32,6)(25,6)
\psline(19,10)(25,6)
\psline(25,6)(19,2)
\rput(28,4){$\scr{|n_1+n_2|}$}
\rput(36,6){$\scr{n_1+n_2}$}
\psline{<-|}(42,1)(35,1)
\psline{<-|}(42,12)(35,12)
\psline{|-<<}(9,1)(2,1)
\psline{|-<<}(9,12)(2,12)
\psline(12,7)(12,5)
\psline(32,7)(32,5)
\rput(6,-1){$j_1$}
\rput(6,10){$j_2$}
\rput(38,10){$j_2$}
\rput(38,-1){$j_1$}
\rput(11,12){$\scr{n_2}$}
\rput(11,2){$\scr{n_1}$}
\rput(33,12){$\scr{n_2}$}
\rput(32,1){$\scr{n_1}$}
\end{pspicture}
\end{array}
\end{split}\label{newrec}
 \ee 

which means that we are now recoupling only projected Wigner matrices (and not lines). The two recoupling rules \eqref{oldrec} and \eqref{newrec} differ when we consider two group elements, based at the same link, which are projected one into the maximum and the other into the minimum magnetic components. The new recoupling theory can be obtained directly from \eqref{su2rec}: when we replace $SU(2)$ with $U(1)$ group elements into triangles, the group element on the right-hand side does not depend on the spin $k$ and the sum just provides a resolution of the identity leading to \eqref{newrec}.
The formula \eqref{newrec} implies then that the reduction commutes with the product operation on the spinnetwork states, namely if we denote as $\psi_1(g_l)$ $\psi_2(g_l)$ two gauge invariant spinnetworks (having their own intertwiners but the same group element associated with the link $l$)  and with $P$ the projection operator, we have $P[\psi_1(g_l)] \;\; P[\psi_2(g_l)]=P[\psi_1(g_l)\psi_2(g_l)]$. Note that this property does not hold with \eqref{oldrec}.

In the following we will outline how the new choice gives us an expression of the scalar constraint operator matrix elements, which simplifies the analysis of the semiclassical limit, allowing us to consider the generic case of six-valent nodes.

\section{Action of the scalar constraint operator}\label{IV}
As discussed in \cite{Alesci:2014uha}, we consider only that part of the scalar constraint whose Poisson action preserves the diagonal gauge condition for the metric. 
This means that we are neglecting the following contributions: i) the scalar curvature of the spatial manifold, ii) non-local terms which arise because of the gauge-fixing conditions. Both these terms can be neglected close to homogeneity and in the limit in which the BKL condition holds for the inhomogeneous extension of Bianchi I model.  

Within this scheme, the scalar constraint is proportional to the Euclidean part of the full scalar constraint $H_E$, and we quantize it by taking the expression of LQG \cite{Thiemann:1996aw,Gaul:2000ba,io e antonia,Alesci:2013kpa} (regularized via a cubulation of the spatial manifold) and replacing the operators in $\mathcal{H}$ with those in ${}^{R}\mathcal{H}$. 

Hence, we study the following Hamiltonian 
\be\label{hamEc}
{}^{R}\hat{H}=\frac{1}{16\pi G\gamma^2}\sum_{\vgraph} {}^{R}\hat{H}^\vgraph_{E}[N]=\frac{1}{16\pi G\gamma^2}\sum_{\vgraph}\sum_{\cube} {}^{R}\hat{H}^\vgraph_{E\cube}[N], 
\ee
where the summations extend over all the nodes $\vgraph$ of the graph the state is based at and over all the triples of links emanating from the same node, while the operator ${}^{R}\hat{H}^\vgraph_{E\cube}[N]$ acting on n-valent nodes is given by
\be
{}^{R}\hat{H}^\vgraph_{E\cube}[N]:= \frac{4i}{8\pi \gamma l^2_P\,c(n)}\mathcal{N}(\vgraph)  \, \epsilon^{ijk} \,
   \mathrm{Tr}\Big[{}^{R}\hat{h}_{\alpha_{ij}} {}^{R}\hat{h}^{-1}_{s_{k}} \big[{}^R\hat{h}_{s_{k}},{}^{R}\hat{V}\big]\Big]\qquad c(n)=2^{n-3}\,, 
   \label{Hridotto}
\ee

$\mathcal{N}(\vgraph)$ being the lapse function in $\vgraph$, while the holonomies are in the fundamental representation.
The trace now means summing over the two possible projections into ${}^{R}\mathcal{H}$ we can perform: on the maximum ($\mu=\frac{1}{2}$) and on the minimum ($\mu=-\frac{1}{2}$) magnetic numbers. Note the use of a different normalization factor with respect to \cite{Thiemann:1996aw,Gaul:2000ba,io e antonia,Alesci:2013kpa}, due to a regularization based on a cubulation instead of a triangulation (there is a factor 6 between the volumes of a parallelepiped and of a tetrahedron sharing three converging edges). The coefficient $c(n)$ is the total number of non-coplanar triples.

It is worth noting how the volume operator ${}^{R}\hat{V}$ is diagonal, such that the matrix elements of ${}^{R}\hat{\mathcal{H}}$ in the basis \eqref{base finale} can be analytically determined.   

Let us first consider the three-valent node discussed in \cite{Alesci:2014uha}, {\it i.e.}  
\be
|\vgraph^z\rangle_R=
\ba
\ifx\JPicScale\undefined\def\JPicScale{0.6}\fi
\psset{unit=\JPicScale mm}
\psset{linewidth=0.3,dotsep=1,hatchwidth=0.3,hatchsep=1.5,shadowsize=1,dimen=middle}
\psset{dotsize=0.7 2.5,dotscale=1 1,fillcolor=black}
\psset{arrowsize=1 2,arrowlength=1,arrowinset=0.25,tbarsize=0.7 5,bracketlength=0.15,rbracketlength=0.15}
\begin{pspicture}(0,0)(101,143)
\psline{|*-}(19,78)(21.83,80.83)
\rput(17,81){$\scr{j_x}$}
\rput{45}(28.5,87.5){\psellipse[](0,0)(2.21,-2.12)}
\rput(24,91){$\scr{h_{x}}$}
\psline[fillstyle=solid]{-|}(42.83,101.83)(40,99)
\pspolygon[](46,103)(44,101)(42,103)(44,105)
\psline[fillstyle=solid](51,110)(44.84,103.84)
\rput(40,104){$\scr{R_x}$}
\psline[fillstyle=solid]{|*-}(16.83,75.83)(14,73)
\rput(10,74){$\scr{R^{-1}_x}$}
\rput(21,85){$\scr{R_x}$}
\psline[fillstyle=solid](26.83,85.83)(24,83)
\psline(30,89)(32.82,91.82)
\pspolygon[](36,93)(34,91)(32,93)(34,95)
\rput(32,97){$\scr{R^{-1}_x}$}
\psline[fillstyle=solid]{|-}(38,97)(35.17,94.17)
\rput(11,77){$\scr{j_x}$}
\psline{|*-}(51,137)(51,133)
\rput(44,129){$\scr{j_z}$}
\rput{0}(51,130){\psellipse[](0,0)(3,-3)}
\rput(51,130){$\scr{h_{z}}$}
\psline[fillstyle=solid]{-|}(51,110)(51,120)
\psline[fillstyle=solid]{|*-}(51,139)(51,143)
\psline{-|}(51,127)(51,123)
\rput(46,141){$\scr{j_z}$}
\pspolygon[](25,82)(23,80)(21,82)(23,84)
\psline[fillstyle=solid](12,71)(6,65)
\pspolygon[](15,72)(13,70)(11,72)(13,74)
\psline{|*-}(83,78)(80.17,80.83)
\rput(84,82){$\scr{j_y}$}
\rput(76,91){$\scr{h_{y}}$}
\psline[fillstyle=solid]{-|}(59.17,101.83)(62,99)
\pspolygon[](58,105)(60,103)(58,101)(56,103)
\rput(62,104){$\scr{R_y}$}
\psline[fillstyle=solid]{|*-}(85.17,75.83)(88,73)
\rput(95,73){$\scr{R^{-1}_y}$}
\rput(79,86){$\scr{R_y}$}
\psline[fillstyle=solid](75.17,85.83)(78,83)
\psline(72,89)(69.18,91.82)
\pspolygon[](68,95)(70,93)(68,91)(66,93)
\rput(72,96){$\scr{R^{-1}_y}$}
\psline[fillstyle=solid]{|-}(64,97)(66.83,94.17)
\rput(90,77){$\scr{j_y}$}
\pspolygon[](79,84)(81,82)(79,80)(77,82)
\psline[fillstyle=solid](90,71)(96,65)
\pspolygon[](89,74)(91,72)(89,70)(87,72)
\rput{43.83}(73.44,87.44){\psellipse[](0,0)(2.21,-2.12)}
\rput(14,60){$\scr{j_l}$}
\rput(80,64){$\scr{j_l}$}
\psline[border=0.3,fillstyle=solid](51,110)(57.16,103.84)
\rput(47,120){$\scr{j_z}$}
\psline{|*-}(42,39)(39.17,41.83)
\rput(43,43){$\scr{j_l}$}
\rput(35,52){$\scr{h_{l_y}}$}
\psline[fillstyle=solid]{-|}(18.17,62.83)(21,60)
\pspolygon[](17,66)(19,64)(17,62)(15,64)
\rput(21,65){$\scr{R_y}$}
\psline[fillstyle=solid]{|*-}(44.17,36.83)(47,34)
\rput(44,30){$\scr{R^{-1}_y}$}
\rput(38,47){$\scr{R_y}$}
\psline[fillstyle=solid](34.17,46.83)(37,44)
\psline(31,50)(28.18,52.82)
\pspolygon[](27,56)(29,54)(27,52)(25,54)
\rput(31,57){$\scr{R^{-1}_y}$}
\psline[fillstyle=solid]{|-}(23,58)(25.83,55.17)
\rput(49,38){$\scr{j_l}$}
\pspolygon[](38,45)(40,43)(38,41)(36,43)
\psline[fillstyle=solid](49,32)(51.83,29.17)
\pspolygon[](48,35)(50,33)(48,31)(46,33)
\rput{43.83}(32.44,48.44){\psellipse[](0,0)(2.21,-2.12)}
\psline[fillstyle=solid](11,70)(16.16,64.84)
\psline{|*-}(80,57)(77.17,54.17)
\rput(64,38){$\scr{j_l}$}
\rput{45}(70.5,47.5){\psellipse[](0,0)(2.21,-2.12)}
\rput(75,46){$\scr{h^{-1}_{l_x}}$}
\psline[fillstyle=solid]{-|}(56.17,33.17)(59,36)
\pspolygon[](53,32)(55,34)(57,32)(55,30)
\psline[fillstyle=solid](52,29)(54.16,31.16)
\rput(58,28){$\scr{R_x}$}
\psline[fillstyle=solid]{|*-}(82.17,59.17)(85,62)
\rput(88,58){$\scr{R^{-1}_x}$}
\rput(80,52){$\scr{R_x}$}
\psline[fillstyle=solid](72.17,49.17)(75,52)
\psline(69,46)(66.18,43.18)
\pspolygon[](63,42)(65,44)(67,42)(65,40)
\rput(71,40){$\scr{R^{-1}_x}$}
\psline[fillstyle=solid]{|-}(61,38)(63.83,40.83)
\pspolygon[](74,53)(76,55)(78,53)(76,51)
\psline[fillstyle=solid](87,64)(92,69)
\pspolygon[](84,63)(86,65)(88,63)(86,61)
\rput(41,108){$\scr{j_x}$}
\rput(58,108){$\scr{j_y}$}
\rput(101,67){$\scr{j'_y}$}
\rput(3,68){$\scr{j'_x}$}
\end{pspicture}
\ea
\label{v3}
\ee
where we called the three fiducial direction $x,y,z$ and we put a loop in the plane $xy$.
For simplicity, we assumed that the Wigner matrices in the state $|\vgraph^z\rangle_R$ have all been projected on the maximum magnetic number, thus $m_l=j_l$. 

The operator \eqref{hamEc} will generically attach loops in all the planes, but since we are considering a non-graph changing operator we are going to discuss only the case in which it provides a loop overlapping the one already present in $|\vgraph^z\rangle_R$.

The action of ${}^{R}\hat{\mathcal{H}}$ coincides with the one discussed in \cite{Alesci:2014uha}, until we have to recouple reduced basis elements and use the new recoupling rule \eqref{newrec}. Hence, looking only at the central node we have
\be
\begin{split}
&\mathrm{Tr}\Big[{}^{R}\hat{h}_{\alpha_{[xy]}} {}^{R}\hat{h}^{-1}_{s_{z}} {}^{R}\!\hat{V} {}^R\hat{h}_{s_{z}}\Big] |\vgraph^z\rangle_R=\\
&=
(8\pi\gamma l_P^2)^{3/2}\sum_{\mu=\pm \frac{1}{2}} \sqrt{j_x\;j_y\;(j_z+\mu)}\; s(\mu)  C^{1\,0}_{\frac{1}{2}\,\frac{1}{2} \;\frac{1}{2}\, -\frac{1}{2}}
\ba
\ifx\JPicScale\undefined\def\JPicScale{0.7}\fi
\psset{unit=\JPicScale mm}
\psset{linewidth=0.3,dotsep=1,hatchwidth=0.3,hatchsep=1.5,shadowsize=1,dimen=middle}
\psset{dotsize=0.7 2.5,dotscale=1 1,fillcolor=black}
\psset{arrowsize=1 2,arrowlength=1,arrowinset=0.25,tbarsize=0.7 5,bracketlength=0.15,rbracketlength=0.15}
\begin{pspicture}(0,50)(95,143)
\psline{|*-}(19,78)(21.83,80.83)
\rput(17,81){$\scr{j_x}$}
\rput{45}(28.5,87.5){\psellipse[](0,0)(2.21,-2.12)}
\rput(24,91){$\scr{h_{x}}$}
\psline[fillstyle=solid]{-|}(42.83,101.83)(40,99)
\pspolygon[](46,103)(44,101)(42,103)(44,105)
\rput(40,104){$\scr{R_x}$}
\psline[fillstyle=solid]{|*-}(16.83,75.83)(14,73)
\rput(10,74){$\scr{R^{-1}_x}$}
\rput(21,85){$\scr{R_x}$}
\psline[fillstyle=solid](26.83,85.83)(24,83)
\psline(30,89)(32.82,91.82)
\pspolygon[](36,93)(34,91)(32,93)(34,95)
\rput(32,97){$\scr{R^{-1}_x}$}
\psline[fillstyle=solid]{|-}(38,97)(35.17,94.17)
\rput(43,108){$\scr{j_x}$}
\rput(13,77){$\scr{j_x}$}
\psline{|*-}(51,137)(51,133)
\rput(44,129){$\scr{j_z}$}
\rput{0}(51,130){\psellipse[](0,0)(3,-3)}
\rput(51,130){$\scr{h_{z}}$}
\psline[fillstyle=solid]{-|}(51,110)(51,120)
\psline[fillstyle=solid]{|*-}(51,139)(51,143)
\psline{-|}(51,127)(51,123)
\rput(46,141){$\scr{j_z}$}
\pspolygon[](25,82)(23,80)(21,82)(23,84)
\psline[fillstyle=solid](12,71)(9.17,68.17)
\pspolygon[](15,72)(13,70)(11,72)(13,74)
\psline{|*-}(83,78)(80.17,80.83)
\rput(84,82){$\scr{j_y}$}
\rput(76,91){$\scr{h_{y}}$}
\psline[fillstyle=solid]{-|}(59.17,101.83)(62,99)
\rput(62,104){$\scr{R_y}$}
\psline[fillstyle=solid]{|*-}(85.17,75.83)(88,73)
\rput(95,73){$\scr{R^{-1}_y}$}
\rput(79,86){$\scr{R_y}$}
\psline[fillstyle=solid](75.17,85.83)(78,83)
\psline(72,89)(69.18,91.82)
\pspolygon[](68,95)(70,93)(68,91)(66,93)
\rput(72,96){$\scr{R^{-1}_y}$}
\psline[fillstyle=solid]{|-}(64,97)(66.83,94.17)
\rput(57,107){$\scr{j_y}$}
\rput(90,77){$\scr{j_y}$}
\pspolygon[](79,84)(81,82)(79,80)(77,82)
\psline[fillstyle=solid](90,71)(92.83,68.17)
\pspolygon[](89,74)(91,72)(89,70)(87,72)
\rput{43.83}(73.44,87.44){\psellipse[](0,0)(2.21,-2.12)}
\psline{|*-}(42,93)(39.17,90.17)
\rput(26,74){$\scr{\frac{1}{2}}$}
\rput{45}(32.5,83.5){\psellipse[](0,0)(2.21,-2.12)}
\rput(37,84){$\scr{h_{x}}$}
\psline[fillstyle=solid]{-|}(18.17,69.17)(21,72)
\pspolygon[](15,68)(17,70)(19,68)(17,66)
\psline[fillstyle=solid](14,65)(16.16,67.16)
\rput(21,67){$\scr{R^{-1}_x}$}
\psline[fillstyle=solid]{|*-}(44.17,95.17)(47,98)
\rput(48,94){$\scr{R_x}$}
\rput(42,88){$\scr{R^{-1}_x}$}
\psline[fillstyle=solid](34.17,85.17)(37,88)
\psline(31,82)(28.18,79.18)
\pspolygon[](25,78)(27,80)(29,78)(27,76)
\rput(31,78){$\scr{R_x}$}
\psline[fillstyle=solid]{|-}(23,74)(25.83,76.83)
\rput(18,63){$\scr{\frac{1}{2}}$}
\rput(51,104){$\scr{\frac{1}{2}}$}
\pspolygon[](36,89)(38,91)(40,89)(38,87)
\psline[fillstyle=solid](49,100)(53,104)
\pspolygon[](46,99)(48,101)(50,99)(48,97)
\psline{|*-}(60,93)(62.83,90.17)
\rput(74,74){$\scr{\frac{1}{2}}$}
\rput(63,82){$\scr{h^{-1}_{y}}$}
\psline[fillstyle=solid]{-|}(83.83,69.17)(81,72)
\pspolygon[](85,66)(83,68)(85,70)(87,68)
\psline[fillstyle=solid](89,64)(85.84,67.16)
\rput(81,67){$\scr{R_y}$}
\psline[fillstyle=solid]{|*-}(57.83,95.17)(55,98)
\rput(55,94){$\scr{R_y^{-1}}$}
\rput(60,88){$\scr{R_y}$}
\psline[fillstyle=solid](67.83,85.17)(65,88)
\psline(71,82)(73.82,79.18)
\pspolygon[](75,76)(73,78)(75,80)(77,78)
\rput(70,77){$\scr{R^{-1}_y}$}
\psline[fillstyle=solid]{|-}(79,74)(76.17,76.83)
\rput(86,64){$\scr{\frac{1}{2}}$}
\rput(60,111){$\scr{1}$}
\pspolygon[](64,87)(62,89)(64,91)(66,89)
\pspolygon[](54,97)(52,99)(54,101)(56,99)
\rput{45}(69.56,83.56){\psellipse[](0,0)(2.21,-2.12)}
\psline[fillstyle=solid](53,104)(61,115)
\psline[fillstyle=solid](54,101)(53,104)
\rput(49,112){$\scr{j_z}$}
\rput(47,120){$\scr{j_z}$}
\psline[border=0.3,fillstyle=solid](51,110)(57,104)
\psline[fillstyle=solid](51,110)(44.84,103.84)
\pspolygon[](58,105)(60,103)(58,101)(56,103)
\rput(62,116){$\scr{0}$}
\end{pspicture}
\ea
\end{split}.
\ee

If one now recouples the U(1) group elements $h_x$ and $h_y^{-1}$, one ends up with the following expression
\be
\begin{split}
&\mathrm{Tr}\Big[{}^{R}\hat{h}_{\alpha_{[xy]}} {}^{R}\hat{h}^{-1}_{s_{z}} {}^{R}\!\hat{V} {}^R\hat{h}_{s_{z}}\Big] |\vgraph^z\rangle_R
=\\
&=(8\pi\gamma l_P^2)^{3/2}\sum_{\mu_x,\mu_y=\pm \frac{1}{2}}\sum_{\mu=\pm \frac{1}{2}} \sqrt{j_x\;j_y\;(j_z+\mu)}\; s(\mu)  C^{1\,0}_{\frac{1}{2}\,\frac{1}{2} \; \frac{1}{2}\,-\frac{1}{2}}
\ba
\ifx\JPicScale\undefined\def\JPicScale{0.7}\fi
\psset{unit=\JPicScale mm}
\psset{linewidth=0.3,dotsep=1,hatchwidth=0.3,hatchsep=1.5,shadowsize=1,dimen=middle}
\psset{dotsize=0.7 2.5,dotscale=1 1,fillcolor=black}
\psset{arrowsize=1 2,arrowlength=1,arrowinset=0.25,tbarsize=0.7 5,bracketlength=0.15,rbracketlength=0.15}
\begin{pspicture}(0,0)(95,143)
\psline[fillstyle=solid]{-|}(42.83,101.83)(40,99)
\pspolygon[](46,103)(44,101)(42,103)(44,105)
\rput(40,104){$\scr{R_x}$}
\psline[fillstyle=solid]{|*-}(16.83,75.83)(14,73)
\rput(10,74){$\scr{R^{-1}_x}$}
\rput(43,108){$\scr{j_x}$}
\rput(13,77){$\scr{j_x}$}
\psline{|*-}(51,137)(51,133)
\rput(44,129){$\scr{j_z}$}
\rput{0}(51,130){\psellipse[](0,0)(3,-3)}
\rput(51,130){$\scr{h_{z}}$}
\psline[fillstyle=solid]{-|}(51,110)(51,120)
\psline[fillstyle=solid]{|*-}(51,139)(51,143)
\psline{-|}(51,127)(51,123)
\rput(46,141){$\scr{j_z}$}
\psline[fillstyle=solid](12,71)(9.17,68.17)
\pspolygon[](15,72)(13,70)(11,72)(13,74)
\psline{|*-}(81,76)(78.17,78.83)
\rput(66,82){$\scr{j_y-\mu_y}$}
\rput(74,89){$\scr{h_{y}}$}
\psline[fillstyle=solid]{-|}(59.17,101.83)(62,99)
\rput(62,104){$\scr{R_y}$}
\psline[fillstyle=solid]{|*-}(85.17,75.83)(88,73)
\rput(95,73){$\scr{R^{-1}_y}$}
\rput(77,84){$\scr{R_y}$}
\psline[fillstyle=solid](73.17,83.83)(76,81)
\psline(70,87)(67.18,89.82)
\pspolygon[](66,93)(68,91)(66,89)(64,91)
\rput(70,94){$\scr{R^{-1}_y}$}
\psline[fillstyle=solid]{|-}(62,95)(64.83,92.17)
\rput(57,107){$\scr{j_y}$}
\rput(90,77){$\scr{j_y}$}
\pspolygon[](77,82)(79,80)(77,78)(75,80)
\psline[fillstyle=solid](90,71)(92.83,68.17)
\pspolygon[](89,74)(91,72)(89,70)(87,72)
\rput{43.83}(71.44,85.44){\psellipse[](0,0)(2.21,-2.12)}
\psline{|*-}(40,95)(37.17,92.17)
\rput{45}(30.5,85.5){\psellipse[](0,0)(2.21,-2.12)}
\rput(35,86){$\scr{h_{x}}$}
\psline[fillstyle=solid]{-|}(18.17,69.17)(21,72)
\pspolygon[](15,68)(17,70)(19,68)(17,66)
\psline[fillstyle=solid](14,65)(16.16,67.16)
\rput(21,67){$\scr{R^{-1}_x}$}
\psline[fillstyle=solid]{|*-}(44.17,95.17)(47,98)
\rput(48,94){$\scr{R_x}$}
\rput(40,90){$\scr{R^{-1}_x}$}
\psline[fillstyle=solid](32.17,87.17)(35,90)
\psline(29,84)(26.18,81.18)
\pspolygon[](23,80)(25,82)(27,80)(25,78)
\rput(29,80){$\scr{R_x}$}
\psline[fillstyle=solid]{|-}(21,76)(23.83,78.83)
\rput(18,63){$\scr{\frac{1}{2}}$}
\rput(51,104){$\scr{\frac{1}{2}}$}
\pspolygon[](34,91)(36,93)(38,91)(36,89)
\psline[fillstyle=solid](49,100)(53,104)
\pspolygon[](46,99)(48,101)(50,99)(48,97)
\psline[fillstyle=solid]{-|}(83.83,69.17)(81,72)
\pspolygon[](85,66)(83,68)(85,70)(87,68)
\psline[fillstyle=solid](89,64)(85.84,67.16)
\rput(81,67){$\scr{R_y}$}
\psline[fillstyle=solid]{|*-}(57.83,95.17)(55,98)
\rput(55,94){$\scr{R_y^{-1}}$}
\rput(86,64){$\scr{\frac{1}{2}}$}
\rput(60,111){$\scr{1}$}
\pspolygon[](54,97)(52,99)(54,101)(56,99)
\psline[fillstyle=solid](53,104)(61,115)
\psline[fillstyle=solid](54,101)(53,104)
\rput(49,112){$\scr{j_z}$}
\rput(47,120){$\scr{j_z}$}
\psline[border=0.3,fillstyle=solid](51,110)(57,104)
\psline[fillstyle=solid](51,110)(44.84,103.84)
\pspolygon[](58,105)(60,103)(58,101)(56,103)
\rput(62,116){$\scr{0}$}
\rput(25,90){$\scr{j_x+\mu_x}$}
\end{pspicture}
\ea
\end{split}
\ee 
It worth noting how the state above does not look like an elements of ${}^{R}\mathcal{H}$, since the coefficient at the central node is not an intertwiner. Indeed, since intertwiners are just numbers, the state above is proportional to a basis element in ${}^{R}\mathcal{H}$ with a factor given by the ratio of the coefficient at the central node and the right intertwiner. One must just check the right intertwiner does not vanish and it is the case here provided that the original intertwiner in \eqref{v3} is nonvanishing (one must just verify that the triangular inequality holds for the involved spin numbers).

By repeating the same analysis for the other (planar) nodes, one ends up with the following expression:

\be
\label{nuova H}
\begin{split}
&\mathrm{Tr}\Big[{}^{R}\hat{h}_{\alpha_{[xy]}} {}^{R}\hat{h}^{-1}_{s_{z}} {}^{R}\!\hat{V} {}^R\hat{h}_{s_{z}}\Big] |\vgraph^z\rangle_R
=\\
&=(8\pi\gamma l_P^2)^{3/2}\sum_{\mu'_x,\mu'_y,\mu_x,\mu_y=\pm \frac{1}{2}}\sum_{\mu=\pm \frac{1}{2}} \sqrt{j_x\;j_y\;(j_z+\mu)}\; s(\mu)  C^{1\,0}_{\frac{1}{2}\,\frac{1}{2} \; \frac{1}{2}\,-\frac{1}{2}}\\
&\ba
\ifx\JPicScale\undefined\def\JPicScale{0.7}\fi
\psset{unit=\JPicScale mm}
\psset{linewidth=0.3,dotsep=1,hatchwidth=0.3,hatchsep=1.5,shadowsize=1,dimen=middle}
\psset{dotsize=0.7 2.5,dotscale=1 1,fillcolor=black}
\psset{arrowsize=1 2,arrowlength=1,arrowinset=0.25,tbarsize=0.7 5,bracketlength=0.15,rbracketlength=0.15}
\begin{pspicture}(0,0)(104,143)
\psline[fillstyle=solid]{-|}(42.83,101.83)(40,99)
\pspolygon[](46,103)(44,101)(42,103)(44,105)
\rput(40,104){$\scr{R_x}$}
\rput(43,108){$\scr{j_x}$}
\rput(13,77){$\scr{j_x}$}
\psline{|*-}(51,137)(51,133)
\rput(44,129){$\scr{j_z}$}
\rput{0}(51,130){\psellipse[](0,0)(3,-3)}
\rput(51,130){$\scr{h_{z}}$}
\psline[fillstyle=solid]{-|}(51,110)(51,120)
\psline[fillstyle=solid]{|*-}(51,139)(51,143)
\psline{-|}(51,127)(51,123)
\rput(46,141){$\scr{j_z}$}
\psline{|*-}(81,76)(78.17,78.83)
\rput(66,82){$\scr{j_y-\mu_y}$}
\rput(74,89){$\scr{h_{y}}$}
\psline[fillstyle=solid]{-|}(59.17,101.83)(62,99)
\rput(62,104){$\scr{R_y}$}
\rput(77,84){$\scr{R_y}$}
\psline[fillstyle=solid](73.17,83.83)(76,81)
\psline(70,87)(67.18,89.82)
\pspolygon[](66,93)(68,91)(66,89)(64,91)
\rput(70,94){$\scr{R^{-1}_y}$}
\psline[fillstyle=solid]{|-}(62,95)(64.83,92.17)
\rput(57,107){$\scr{j_y}$}
\rput(90,77){$\scr{j_y}$}
\pspolygon[](77,82)(79,80)(77,78)(75,80)
\rput{43.83}(71.44,85.44){\psellipse[](0,0)(2.21,-2.12)}
\psline{|*-}(40,95)(37.17,92.17)
\rput{45}(30.5,85.5){\psellipse[](0,0)(2.21,-2.12)}
\rput(35,86){$\scr{h_{x}}$}
\psline[fillstyle=solid]{|*-}(44.17,95.17)(47,98)
\rput(48,94){$\scr{R_x}$}
\rput(40,90){$\scr{R^{-1}_x}$}
\psline[fillstyle=solid](32.17,87.17)(35,90)
\psline(29,84)(26.18,81.18)
\pspolygon[](23,80)(25,82)(27,80)(25,78)
\rput(29,80){$\scr{R_x}$}
\psline[fillstyle=solid]{|-}(21,76)(23.83,78.83)
\rput(51,104){$\scr{\frac{1}{2}}$}
\pspolygon[](34,91)(36,93)(38,91)(36,89)
\psline[fillstyle=solid](49,100)(53,104)
\pspolygon[](46,99)(48,101)(50,99)(48,97)
\psline[fillstyle=solid]{|*-}(57.83,95.17)(55,98)
\rput(55,94){$\scr{R_y^{-1}}$}
\rput(60,111){$\scr{1}$}
\pspolygon[](54,97)(52,99)(54,101)(56,99)
\psline[fillstyle=solid](53,104)(61,115)
\psline[fillstyle=solid](54,101)(53,104)
\rput(49,112){$\scr{j_z}$}
\rput(47,120){$\scr{j_z}$}
\psline[border=0.3,fillstyle=solid](51,110)(57,104)
\psline[fillstyle=solid](51,110)(44.84,103.84)
\pspolygon[](58,105)(60,103)(58,101)(56,103)
\rput(62,116){$\scr{0}$}
\rput(25,90){$\scr{j_x+\mu_x}$}
\psline[fillstyle=solid]{|*-}(16.83,75.83)(14,73)
\rput(10,74){$\scr{R^{-1}_x}$}
\psline[fillstyle=solid](12,71)(-2,57)
\pspolygon[](15,72)(13,70)(11,72)(13,74)
\psline[fillstyle=solid]{|*-}(85.17,75.83)(88,73)
\rput(95,73){$\scr{R^{-1}_y}$}
\psline[fillstyle=solid](90,71)(104,57)
\pspolygon[](89,74)(91,72)(89,70)(87,72)
\rput(11,46){$\scr{j_l}$}
\rput(93,50){$\scr{j_l}$}
\psline[fillstyle=solid]{-|}(18.17,49.83)(21,47)
\pspolygon[](17,53)(19,51)(17,49)(15,51)
\rput(9,50){$\scr{R_y}$}
\psline[fillstyle=solid]{|*-}(44.17,23.83)(47,21)
\rput(44,17){$\scr{R^{-1}_y}$}
\rput(52,12){$\scr{j_l}$}
\psline[fillstyle=solid](49,19)(51.83,16.17)
\pspolygon[](48,22)(50,20)(48,18)(46,20)
\psline[fillstyle=solid](4,63)(16.16,51.84)
\psline[fillstyle=solid]{-|}(56.17,20.17)(59,23)
\pspolygon[](53,19)(55,21)(57,19)(55,17)
\psline[fillstyle=solid](52,16)(54.16,18.16)
\rput(58,15){$\scr{R_x}$}
\psline[fillstyle=solid]{|*-}(82.17,46.17)(85,49)
\rput(88,45){$\scr{R^{-1}_x}$}
\psline[fillstyle=solid](87,51)(99,62)
\pspolygon[](84,50)(86,52)(88,50)(86,48)
\rput(104,62){$\scr{j'_y}$}
\rput(-2,62){$\scr{j'_x}$}
\psline[fillstyle=solid]{|*-}(18.83,70.83)(16,68)
\rput(21,67){$\scr{R^{-1}_x}$}
\rput(25,72){$\scr{\mu_x}$}
\psline[fillstyle=solid](14,66)(13,65)
\pspolygon[](17,67)(15,65)(13,67)(15,69)
\psline[fillstyle=solid]{|*-}(83.17,70.83)(86,68)
\rput(80,68){$\scr{R_y}$}
\rput(78,72){$\scr{\mu_y}$}
\psline[fillstyle=solid](88,66)(90,64)
\pspolygon[](87,69)(89,67)(87,65)(85,67)
\rput(24,59){$\scr{\mu'_y}$}
\rput(77,57){$\scr{\mu'_x}$}
\psline{|*-}(40,33)(37.17,35.83)
\rput(43,35){$\scr{\mu'_y}$}
\rput(36,46){$\scr{h_{l_y}}$}
\psline[fillstyle=solid]{-|}(18.17,59.83)(21,57)
\pspolygon[](17,63)(19,61)(17,59)(15,61)
\rput(21,62){$\scr{R_y}$}
\psline[fillstyle=solid]{|*-}(44.17,33.83)(47,31)
\rput(49,34){$\scr{R^{-1}_y}$}
\rput(36,41){$\scr{R_y}$}
\psline[fillstyle=solid](32.17,40.83)(35,38)
\psline(29,44)(26.18,46.82)
\pspolygon[](25,50)(27,48)(25,46)(23,48)
\rput(29,51){$\scr{R^{-1}_y}$}
\psline[fillstyle=solid]{|-}(21,52)(23.83,49.17)
\pspolygon[](36,39)(38,37)(36,35)(34,37)
\psline[fillstyle=solid](49,29)(51.83,26.17)
\pspolygon[](48,32)(50,30)(48,28)(46,30)
\rput{43.83}(30.44,42.44){\psellipse[](0,0)(2.21,-2.12)}
\psline[fillstyle=solid](13,65)(16.16,61.84)
\psline{|*-}(82,51)(79.17,48.17)
\rput(56,37){$\scr{\mu'_x}$}
\rput{45}(72.5,41.5){\psellipse[](0,0)(2.21,-2.12)}
\rput(66,45){$\scr{h^{-1}_{l_x}}$}
\psline[fillstyle=solid]{-|}(56.17,30.17)(59,33)
\pspolygon[](53,29)(55,31)(57,29)(55,27)
\psline[fillstyle=solid](52,26)(54.16,28.16)
\rput(55,34){$\scr{R_x}$}
\psline[fillstyle=solid]{|*-}(82.17,56.17)(85,59)
\rput(80,62){$\scr{R^{-1}_x}$}
\rput(71,50){$\scr{R_x}$}
\psline[fillstyle=solid](74.17,43.17)(77,46)
\psline(71,40)(68.18,37.18)
\pspolygon[](65,36)(67,38)(69,36)(67,34)
\rput(63,40){$\scr{R^{-1}_x}$}
\psline[fillstyle=solid]{|-}(63,32)(65.83,34.83)
\pspolygon[](76,47)(78,49)(80,47)(78,45)
\psline[fillstyle=solid](87,61)(90,64)
\pspolygon[](84,60)(86,62)(88,60)(86,58)
\rput(24,38){$\scr{j_l+\mu'_y}$}
\rput(77,37){$\scr{j_l+\mu'_x}$}
\end{pspicture}
\ea
\end{split}
\ee

The coefficient left from the intertwiner reduction at the node are just standard $SU(2)$ intertwiners contracted with fixed rotation matrices. They can be recoupled node by node with the original intertwiners, producing a 9j symbol at the central node and 6j symbols at the planar nodes {\it summed} over the whole tower of intermediate representations allowed by the tensor product of the intertwiners. This sum is the main difference with the results of \cite{Alesci:2014uha} where only the term arising from the tensor product with maximum total spin was taken into account.

However, it is convenient to keep the products of single intertwiners as in the previous formula, since it outlines how the action of the Hamiltonian does not depend on the valence of the node, thus it can be easily extended to more-than-three-valent nodes. 
Given a six-valent node in a cuboidal lattice

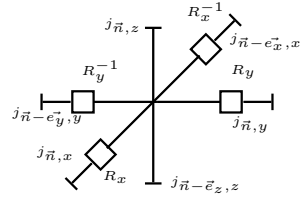
\be
\ba
\ifx\JPicScale\undefined\def\JPicScale{1}\fi
\psset{unit=\JPicScale mm}
\psset{linewidth=0.3,dotsep=1,hatchwidth=0.3,hatchsep=1.5,shadowsize=1,dimen=middle}
\psset{dotsize=0.7 2.5,dotscale=1 1,fillcolor=black}
\psset{arrowsize=1 2,arrowlength=1,arrowinset=0.25,tbarsize=0.7 5,bracketlength=0.15,rbracketlength=0.15}
\begin{pspicture}(50,0)(87,70)
\psline[fillstyle=solid]{-|}(62.83,49.83)(60,47)
\pspolygon[](66,51)(64,49)(62,51)(64,53)
\rput(66,48){$\scr{R_x}$}
\psline[fillstyle=solid]{-|}(71,58)(71,68)
\psline[fillstyle=solid]{-|}(83,58)(87,58)
\rput(83,62){$\scr{R_y}$}
\rput(84,55){$\scr{j_{\vec{n},y}}$}
\rput(67,68){$\scr{j_{\vec{n},z}}$}
\psline[fillstyle=solid](71,58)(64.84,51.84)
\pspolygon[](79.93,59.41)(82.76,59.41)(82.76,56.59)(79.93,56.59)
\psline[fillstyle=solid](71.51,58)(80,58)
\psline{-|}(71,58)(71,47)
\rput(64,62){$\scr{R^{-1}_y}$}
\rput(57,56){$\scr{j_{\vec{n}-\vec{e_y},y}}$}
\psline[fillstyle=solid]{-|}(60,58)(56,58)
\pspolygon[](63.07,56.59)(60.24,56.59)(60.24,59.41)(63.07,59.41)
\psline[fillstyle=solid](71.49,58)(63,58)
\psline[fillstyle=solid]{-|}(79.17,66.17)(82,69)
\pspolygon[](76,65)(78,67)(80,65)(78,63)
\rput(78,70){$\scr{R^{-1}_x}$}
\rput(86,66){$\scr{j_{\vec{n}-\vec{e_x},x}}$}
\psline[fillstyle=solid](71,58)(77.16,64.16)
\rput(78,47){$\scr{j_{\vec{n}-\vec{e}_z,z}}$}
\rput(58,51){$\scr{j_{\vec{n},x}}$}
\end{pspicture}
\ea
\ee
where the position of the node is described by a vector $\vec{n}$, $\vec{e}_i$ with $i=x,y,z$ denotes the vectors along the fiducial directions connecting with the nearest nodes, and the links $j_{\vec{n},i}$ with $i=x,y,z$ are outgoing from the node in $\vec{n}$,
we get the analogous of the expression \eqref{nuova H}

\be
\begin{split}
&\mathrm{Tr}\Big[{}^{R}\hat{h}_{\alpha_{[xy]}} {}^{R}\hat{h}^{-1}_{s_{z}} {}^{R}\!\hat{V} {}^R\hat{h}_{s_{z}}\Big] |\vgraph_{0,0,0}\rangle_R
=\\
&=(8\pi\gamma l_P^2)^{3/2}\sum_{\mu'_x,\mu'_y,\mu_x,\mu_y=\pm \frac{1}{2}}\sum_{\mu=\pm \frac{1}{2}} \sqrt{j_x\;j_y\;(j_z+\mu)}\; s(\mu)  C^{1\,0}_{\frac{1}{2}\,\frac{1}{2} \; \frac{1}{2}\,-\frac{1}{2}}\\
&\ba
\ifx\JPicScale\undefined\def\JPicScale{1}\fi
\psset{unit=\JPicScale mm}
\psset{linewidth=0.3,dotsep=1,hatchwidth=0.3,hatchsep=1.5,shadowsize=1,dimen=middle}
\psset{dotsize=0.7 2.5,dotscale=1 1,fillcolor=black}
\psset{arrowsize=1 2,arrowlength=1,arrowinset=0.25,tbarsize=0.7 5,bracketlength=0.15,rbracketlength=0.15}
\begin{pspicture}(0,0)(173,90)
\psline[fillstyle=solid]{-|}(68.83,70.83)(66,68)
\pspolygon[](72,72)(70,70)(68,72)(70,74)
\rput(66,73){$\scr{R_x}$}
\rput(71,76){$\scr{j_x}$}
\psline[fillstyle=solid]{-|}(77,79)(77,89)
\psline[fillstyle=solid]{-|}(89,79)(93,79)
\rput(89.46,82.54){$\scr{R_y}$}
\rput(83.81,81.12){$\scr{j_y}$}
\rput(74,63){$\scr{R_x}$}
\rput(82,70){$\scr{m}$}
\psline[fillstyle=solid]{|*-}(93,73)(89,73)
\rput(91.59,70.44){$\scr{R_y^{-1}}$}
\rput(81,76){$\scr{1}$}
\pspolygon[](88.76,71.86)(85.93,71.86)(85.93,74.69)(88.76,74.69)
\psline[fillstyle=solid](86,73)(79,73)
\rput(73,89){$\scr{j_z}$}
\psline[fillstyle=solid](77,79)(70.84,72.84)
\pspolygon[](85.93,80.41)(88.76,80.41)(88.76,77.59)(85.93,77.59)
\rput(79,86){$\scr{0}$}
\psline[fillstyle=solid](77.51,79)(86,79)
\psline(77,79)(77,65)
\psline[border=0.3,fillstyle=solid](75,69)(79,73)
\psline[fillstyle=solid]{-|}(66,79)(62,79)
\pspolygon[](69.07,77.59)(66.24,77.59)(66.24,80.41)(69.07,80.41)
\psline[fillstyle=solid](77.49,79)(69,79)
\psline[fillstyle=solid]{-|}(85.17,87.17)(88,90)
\pspolygon[](82,86)(84,88)(86,86)(84,84)
\psline[fillstyle=solid](77,79)(83.16,85.16)
\psline[border=0.3,fillstyle=solid](79,73)(79,84)
\psline[fillstyle=solid]{-|}(18.83,20.83)(16,18)
\pspolygon[](22,22)(20,20)(18,22)(20,24)
\psline[fillstyle=solid]{-|}(27,29)(27,39)
\psline[fillstyle=solid]{-|}(39,29)(43,29)
\rput(39.46,32.54){$\scr{R_y}$}
\rput(21,38){$\scr{j_{\vec{e}_x, z}}$}
\psline[fillstyle=solid](27,29)(20.84,22.84)
\pspolygon[](35.93,30.41)(38.76,30.41)(38.76,27.59)(35.93,27.59)
\psline[fillstyle=solid](27.51,29)(36,29)
\psline(27,29)(27,15)
\psline[fillstyle=solid]{-|}(16,29)(12,29)
\pspolygon[](19.07,27.59)(16.24,27.59)(16.24,30.41)(19.07,30.41)
\psline[fillstyle=solid](27.49,29)(19,29)
\psline[fillstyle=solid]{-|}(35.17,37.17)(38,40)
\pspolygon[](32,36)(34,38)(36,36)(34,34)
\rput(34,41){$\scr{R^{-1}_x}$}
\rput(39,38){$\scr{j_{x}}$}
\psline[fillstyle=solid](27,29)(33.16,35.16)
\psline[fillstyle=solid]{|*-}(70.17,64.17)(73,67)
\pspolygon[](72,68)(74,70)(76,68)(74,66)
\psline{|*-}(66,64)(63.17,61.17)
\rput{45}(56.5,54.5){\psellipse[](0,0)(2.21,-2.12)}
\rput(61,55){$\scr{h_{x}}$}
\rput(66,59){$\scr{R^{-1}_x}$}
\psline[fillstyle=solid](58.17,56.17)(61,59)
\psline(55,53)(52.18,50.18)
\pspolygon[](49,49)(51,51)(53,49)(51,47)
\rput(55,49){$\scr{R_x}$}
\psline[fillstyle=solid]{|-}(47,45)(49.83,47.83)
\pspolygon[](60,60)(62,62)(64,60)(62,58)
\rput(51,59){$\scr{j_x+\mu_x}$}
\psline[fillstyle=solid]{|*-}(46.83,40.83)(44,38)
\rput(48,39){$\scr{R^{-1}_x}$}
\rput(53,42){$\scr{\mu_x}$}
\pspolygon[](45,37)(43,35)(41,37)(43,39)
\psline[fillstyle=solid]{-|}(47.34,33.93)(51.34,33.93)
\pspolygon[](44.27,35.34)(47.1,35.34)(47.1,32.51)(44.27,32.51)
\rput(49,36){$\scr{R_y}$}
\psline[fillstyle=solid](40.03,33.93)(44.5,33.93)
\psline[fillstyle=solid](42,36)(40,34)
\psline{|*-}(123.88,75.88)(119.88,75.88)
\rput(109.03,69.51){$\scr{j_y-\mu_y}$}
\rput(109.74,80.12){$\scr{h_{y}}$}
\rput(115.39,78.71){$\scr{R_y}$}
\psline[fillstyle=solid](112.81,75.88)(116.81,75.88)
\psline(108.32,75.88)(104.33,75.88)
\pspolygon[](101.25,77.29)(104.08,77.29)(104.08,74.46)(101.25,74.46)
\rput(103.37,80.83){$\scr{R^{-1}_y}$}
\psline[fillstyle=solid]{|-}(97.01,75.88)(101.01,75.88)
\pspolygon[](116.81,77.29)(119.64,77.29)(119.64,74.46)(116.81,74.46)
\rput{90}(110.44,75.79){\psellipse[](0,0)(2.21,-2.12)}
\psline{|*-}(83.88,31.88)(79.88,31.88)
\rput(69.74,36.12){$\scr{h_{y}}$}
\rput(75.39,34.71){$\scr{R_y}$}
\psline[fillstyle=solid](72.81,31.88)(76.81,31.88)
\psline(68.32,31.88)(64.33,31.88)
\pspolygon[](61.25,33.29)(64.08,33.29)(64.08,30.46)(61.25,30.46)
\rput(63,36){$\scr{R^{-1}_y}$}
\psline[fillstyle=solid]{|-}(57,32)(61,32)
\pspolygon[](76.81,33.29)(79.64,33.29)(79.64,30.46)(76.81,30.46)
\rput{90}(70.44,31.79){\psellipse[](0,0)(2.21,-2.12)}
\psline[fillstyle=solid]{-|}(148.83,69.83)(146,67)
\pspolygon[](152,71)(150,69)(148,71)(150,73)
\rput(146,72){$\scr{R_x}$}
\rput(151,66){$\scr{j_{\vec{e}_y,x}}$}
\psline[fillstyle=solid]{-|}(157,78)(157,88)
\psline[fillstyle=solid]{-|}(169,78)(173,78)
\psline[fillstyle=solid](157,78)(150.84,71.84)
\pspolygon[](165.93,79.41)(168.76,79.41)(168.76,76.59)(165.93,76.59)
\psline[fillstyle=solid](157.51,78)(166,78)
\psline(157,78)(157,64)
\rput(144,82){$\scr{R^{-1}_y}$}
\rput(151,81){$\scr{j_{y}}$}
\psline[fillstyle=solid]{-|}(146,78)(142,78)
\pspolygon[](149.07,76.59)(146.24,76.59)(146.24,79.41)(149.07,79.41)
\psline[fillstyle=solid](157.49,78)(149,78)
\pspolygon[](162,85)(164,87)(166,85)(164,83)
\psline[fillstyle=solid](157,78)(163.16,84.16)
\psline[fillstyle=solid]{-|}(98.83,20.83)(96,18)
\pspolygon[](102,22)(100,20)(98,22)(100,24)
\psline[fillstyle=solid](107,29)(100.84,22.84)
\psline(107,29)(107,15)
\rput(95,32){$\scr{R^{-1}_y}$}
\psline[fillstyle=solid]{-|}(96,29)(92,29)
\pspolygon[](99.07,27.59)(96.24,27.59)(96.24,30.41)(99.07,30.41)
\psline[fillstyle=solid](107.49,29)(99,29)
\psline[fillstyle=solid]{-|}(115.17,37.17)(118,40)
\pspolygon[](112,36)(114,38)(116,36)(114,34)
\rput(118,35){$\scr{R^{-1}_x}$}
\psline[fillstyle=solid](107,29)(113.16,35.16)
\psline{|*-}(138,64)(135.17,61.17)
\rput{45}(128.5,54.5){\psellipse[](0,0)(2.21,-2.12)}
\rput(133,55){$\scr{h_{x}}$}
\rput(130,61){$\scr{R^{-1}_x}$}
\psline[fillstyle=solid](130.17,56.17)(133,59)
\psline(127,53)(124.18,50.18)
\pspolygon[](121,49)(123,51)(125,49)(123,47)
\rput(117,49){$\scr{R_x}$}
\psline[fillstyle=solid]{|-}(119,45)(121.83,47.83)
\pspolygon[](132,60)(134,62)(136,60)(134,58)
\rput(106,42){$\scr{R_x}$}
\psline[fillstyle=solid]{|*-}(92,35)(96,35)
\rput(93,39){$\scr{R_y^{-1}}$}
\pspolygon[](96.24,36.14)(99.07,36.14)(99.07,33.31)(96.24,33.31)
\psline[fillstyle=solid](99,35)(106,35)
\psline[fillstyle=solid](110,39)(106,35)
\psline[fillstyle=solid]{|*-}(114.83,43.83)(112,41)
\pspolygon[](113,40)(111,38)(109,40)(111,42)
\psline[fillstyle=solid]{|*-}(138.17,67.17)(141,70)
\rput(137,69){$\scr{R^{-1}_x}$}
\pspolygon[](140,71)(142,73)(144,71)(142,69)
\psline[fillstyle=solid]{-|}(137.66,74.07)(133.66,74.07)
\pspolygon[](140.73,72.66)(137.9,72.66)(137.9,75.49)(140.73,75.49)
\rput(136,77){$\scr{R_y}$}
\psline[fillstyle=solid](144.97,74.07)(140.5,74.07)
\psline[fillstyle=solid](143,72)(145,74)
\psline[border=0.3,fillstyle=solid]{-|}(107,29)(107,39)
\psline[fillstyle=solid]{-|}(119,29)(123,29)
\pspolygon[](115.93,30.41)(118.76,30.41)(118.76,27.59)(115.93,27.59)
\psline[fillstyle=solid](107.51,29)(116,29)
\psline[fillstyle=solid]{-|}(165.17,86.17)(168,89)
\rput(33,26){$\scr{j_{\vec{e}_x, y}}$}
\rput(69,27){$\scr{j_{\vec{e}_x, y}+\mu'_y}$}
\rput(93,26){$\scr{j_{\vec{e}_x, y}}$}
\rput(122,38){$\scr{j_{\vec{e}_y,x}}$}
\rput(116,54){$\scr{j_{\vec{e}_y,x}-\mu'_x}$}
\rput(70,62){$\scr{\mu_x}$}
\rput(53,34){$\scr{\mu'_y}$}
\rput(90,35){$\scr{\mu'_y}$}
\rput(114,46){$\scr{\mu'_x}$}
\rput(136,67){$\scr{\mu'_x}$}
\rput(131,74){$\scr{\mu_y}$}
\rput(94,72){$\scr{\mu_y}$}
\end{pspicture}
\ea\,,
\end{split}
\ee
where we consider just one node in the position $\vec{n}=(0,0,0)$. This result has then to be summed over all the possible permutations of the links within the chosen triple.

\section{Semiclassical limit of the Hamiltonian}\label{Scal}
The definition of coherent states in QRLG can be performed according with the tools adopted in full LQG \cite{Thiemann:2000bw,Thiemann:2002vj}. In particular, we define semiclassical states based at a graph $\Gamma$ as in \cite{Alesci:2014uha}: 
\begin{equation}
\psi^{{\bf\alpha}}_{\Gamma{\bf H'}}=\sum_{{\bf m_{l}}}\prod_{n\in\Gamma} \langle{\bf j_{l}}, {\bf x}_n|{\bf m_{l}},  \vec{{\bf u}}_l \rangle^*\;\prod_{l\in\Gamma} \psi^\alpha_{H'_{l}}(m_{l})\;\langle h|\Gamma, {\bf m_l, x_n \bf}\rangle\,,
\label{semiclassici ridotti inv}
\end{equation}
where the functions $\psi^\alpha_{H'_{l}}(m_{l})$ are determined by $H'_l=h_le^{\frac{\alpha}{8\pi\gamma l_P^2}E'_l\tau_l}$ belonging to the complexification of the SU(2) group and giving the classical values for the holonomy $h_l$ and the dual flux $E_l$ around which the state is peaked. The definition of these classical quantities involves the physical varialbes $c_i$ and $p_i$ for $\ell_0=\epsilon$, $\epsilon$ being the coordinate length of the link $l$, while the coordinate area of the dual surface is $\epsilon^2$. The explicit expression of $\psi^\alpha_{H'}(m_{l})$ reads 
\begin{equation}
\psi^\alpha_{H'}(m_{l})=(2j_{l}+1)e^{-j_{l}(j_{l}+1)\frac{\alpha}{2}}e^{i\theta_lm_{l}}e^{\frac{\alpha}{8\pi\gamma l_P^2}E'_im_{l}}\,,\quad j_l=|m_l|\,.
\end{equation}
The insertion of such functions insures good peakedness properties in the limit $\frac{E'}{8\pi\gamma l_P^2}\gg1$, since they approach gaussian functions, {\it i.e.}
\be
\psi^\alpha_{H'}(m_{l})\sim(2j_{l}+1)e^{-\frac{\alpha}{2}\left(m_{l}-\frac{E'_i}{8\pi\gamma l_P^2}\right)^2}e^{i\theta_lm_{l}}\,,\label{simgauss}
\ee
times some phase terms which are representations of the classical holonomy ${}^{l}\!D^{j_l}_{m_lm_l}(h_l)=e^{i\theta_l m_l}$. For a constant physical connection \footnote{``Physical'' means rescaled by the length of the fiducial cell, such that it does not depend on the choice of the fiducial metric. This is the kind of variables adopted in LQC.} $c_l$ along the link $l$, the parameter $\theta_l$ equals $\pm c_l$ and the sign depends on the relative orientation between the link and fiducial directions. Similarly, the physical momenta $p^l$ equals the flux $E'_l=8\pi\gamma l_P^2 \bar{m}_l$, $\bar{m}_l$ being the center of the gaussian function in \eqref{simgauss}. We can now construct coherent states associated with the three-valent dressed node \eqref{v3}, which reads as follows
\be
|\Psi_{H} \; \vgraph^z\rangle=\sum_{j_x,j_y,j_z, j_l}
\ba
\ifx\JPicScale\undefined\def\JPicScale{0.6}\fi
\psset{unit=\JPicScale mm}
\psset{linewidth=0.3,dotsep=1,hatchwidth=0.3,hatchsep=1.5,shadowsize=1,dimen=middle}
\psset{dotsize=0.7 2.5,dotscale=1 1,fillcolor=black}
\psset{arrowsize=1 2,arrowlength=1,arrowinset=0.25,tbarsize=0.7 5,bracketlength=0.15,rbracketlength=0.15}
\begin{pspicture}(0,0)(101,143)
\rput(28,87){$\Psi_{H_{l_x}}(j_x)$}
\psline[fillstyle=solid]{-|}(42.83,101.83)(40,99)
\pspolygon[](46,103)(44,101)(42,103)(44,105)
\psline[fillstyle=solid](51,110)(44.84,103.84)
\rput(40,104){$\scr{R_x}$}
\psline[fillstyle=solid]{|*-}(16.83,75.83)(14,73)
\rput(10,74){$\scr{R^{-1}_x}$}
\rput(11,77){$\scr{j_x}$}
\psline[fillstyle=solid]{-|}(51,110)(51,120)
\psline[fillstyle=solid]{|*-}(51,139)(51,143)
\rput(46,141){$\scr{j_z}$}
\psline[fillstyle=solid](12,71)(6,65)
\pspolygon[](15,72)(13,70)(11,72)(13,74)
\psline[fillstyle=solid]{-|}(59.17,101.83)(62,99)
\pspolygon[](58,105)(60,103)(58,101)(56,103)
\rput(62,104){$\scr{R_y}$}
\psline[fillstyle=solid]{|*-}(85.17,75.83)(88,73)
\rput(95,73){$\scr{R^{-1}_y}$}
\rput(90,77){$\scr{j_y}$}
\psline[fillstyle=solid](90,71)(96,65)
\pspolygon[](89,74)(91,72)(89,70)(87,72)
\rput(14,60){$\scr{j_l}$}
\rput(80,64){$\scr{j_l}$}
\psline[border=0.3,fillstyle=solid](51,110)(57.16,103.84)
\rput(47,120){$\scr{j_z}$}
\psline[fillstyle=solid]{-|}(18.17,62.83)(21,60)
\pspolygon[](17,66)(19,64)(17,62)(15,64)
\rput(21,65){$\scr{R_y}$}
\psline[fillstyle=solid]{|*-}(44.17,36.83)(47,34)
\rput(44,30){$\scr{R^{-1}_y}$}
\rput(49,38){$\scr{j_l}$}
\psline[fillstyle=solid](49,32)(51.83,29.17)
\pspolygon[](48,35)(50,33)(48,31)(46,33)
\psline[fillstyle=solid](11,70)(16.16,64.84)
\psline[fillstyle=solid]{-|}(56.17,33.17)(59,36)
\pspolygon[](53,32)(55,34)(57,32)(55,30)
\psline[fillstyle=solid](52,29)(54.16,31.16)
\rput(58,28){$\scr{R_x}$}
\psline[fillstyle=solid]{|*-}(82.17,59.17)(85,62)
\psline[fillstyle=solid](87,64)(92,69)
\pspolygon[](84,63)(86,65)(88,63)(86,61)
\rput(41,108){$\scr{j_x}$}
\rput(58,108){$\scr{j_y}$}
\rput(101,67){$\scr{j'_y}$}
\rput(3,68){$\scr{j'_x}$}
\rput(73,87){$\Psi_{H_{l_y}}(j_y)$}
\rput(33,48){$\Psi_{H_{l_{l_{y}}}}(j_l)$}
\rput(71,48){$\Psi_{H_{l_{l_{x}}}}(j_l)$}
\rput(51,130){$\Psi_{H_{l_z}}(j_z)$}
\rput(54,113){{\large{*}}}
\rput(84,68){{\large{*}}}
\rput(20,70){{\large{*}}}
\rput(52,25){{\large{*}}}
\end{pspicture}
\ea
\quad
\Bigg|
\ba
\ifx\JPicScale\undefined\def\JPicScale{0.55}\fi
\psset{unit=\JPicScale mm}
\psset{linewidth=0.3,dotsep=1,hatchwidth=0.3,hatchsep=1.5,shadowsize=1,dimen=middle}
\psset{dotsize=0.7 2.5,dotscale=1 1,fillcolor=black}
\psset{arrowsize=1 2,arrowlength=1,arrowinset=0.25,tbarsize=0.7 5,bracketlength=0.15,rbracketlength=0.15}
\begin{pspicture}(0,0)(101,143)
\psline{|*-}(19,78)(21.83,80.83)
\rput(17,81){$\scr{j_x}$}
\rput{45}(28.5,87.5){\psellipse[](0,0)(2.21,-2.12)}
\rput(24,91){$\scr{h_{x}}$}
\psline[fillstyle=solid]{-|}(42.83,101.83)(40,99)
\pspolygon[](46,103)(44,101)(42,103)(44,105)
\psline[fillstyle=solid](51,110)(44.84,103.84)
\rput(40,104){$\scr{R_x}$}
\psline[fillstyle=solid]{|*-}(16.83,75.83)(14,73)
\rput(10,74){$\scr{R^{-1}_x}$}
\rput(21,85){$\scr{R_x}$}
\psline[fillstyle=solid](26.83,85.83)(24,83)
\psline(30,89)(32.82,91.82)
\pspolygon[](36,93)(34,91)(32,93)(34,95)
\rput(32,97){$\scr{R^{-1}_x}$}
\psline[fillstyle=solid]{|-}(38,97)(35.17,94.17)
\rput(11,77){$\scr{j_x}$}
\psline{|*-}(51,137)(51,133)
\rput(44,129){$\scr{j_z}$}
\rput{0}(51,130){\psellipse[](0,0)(3,-3)}
\rput(51,130){$\scr{h_{z}}$}
\psline[fillstyle=solid]{-|}(51,110)(51,120)
\psline[fillstyle=solid]{|*-}(51,139)(51,143)
\psline{-|}(51,127)(51,123)
\rput(46,141){$\scr{j_z}$}
\pspolygon[](25,82)(23,80)(21,82)(23,84)
\psline[fillstyle=solid](12,71)(6,65)
\pspolygon[](15,72)(13,70)(11,72)(13,74)
\psline{|*-}(83,78)(80.17,80.83)
\rput(84,82){$\scr{j_y}$}
\rput(76,91){$\scr{h_{y}}$}
\psline[fillstyle=solid]{-|}(59.17,101.83)(62,99)
\pspolygon[](58,105)(60,103)(58,101)(56,103)
\rput(62,104){$\scr{R_y}$}
\psline[fillstyle=solid]{|*-}(85.17,75.83)(88,73)
\rput(95,73){$\scr{R^{-1}_y}$}
\rput(79,86){$\scr{R_y}$}
\psline[fillstyle=solid](75.17,85.83)(78,83)
\psline(72,89)(69.18,91.82)
\pspolygon[](68,95)(70,93)(68,91)(66,93)
\rput(72,96){$\scr{R^{-1}_y}$}
\psline[fillstyle=solid]{|-}(64,97)(66.83,94.17)
\rput(90,77){$\scr{j_y}$}
\pspolygon[](79,84)(81,82)(79,80)(77,82)
\psline[fillstyle=solid](90,71)(96,65)
\pspolygon[](89,74)(91,72)(89,70)(87,72)
\rput{43.83}(73.44,87.44){\psellipse[](0,0)(2.21,-2.12)}
\rput(14,60){$\scr{j_l}$}
\rput(80,64){$\scr{j_l}$}
\psline[border=0.3,fillstyle=solid](51,110)(57.16,103.84)
\rput(47,120){$\scr{j_z}$}
\psline{|*-}(42,39)(39.17,41.83)
\rput(43,43){$\scr{j_l}$}
\rput(35,52){$\scr{h_{l_y}}$}
\psline[fillstyle=solid]{-|}(18.17,62.83)(21,60)
\pspolygon[](17,66)(19,64)(17,62)(15,64)
\rput(21,65){$\scr{R_y}$}
\psline[fillstyle=solid]{|*-}(44.17,36.83)(47,34)
\rput(44,30){$\scr{R^{-1}_y}$}
\rput(38,47){$\scr{R_y}$}
\psline[fillstyle=solid](34.17,46.83)(37,44)
\psline(31,50)(28.18,52.82)
\pspolygon[](27,56)(29,54)(27,52)(25,54)
\rput(31,57){$\scr{R^{-1}_y}$}
\psline[fillstyle=solid]{|-}(23,58)(25.83,55.17)
\rput(49,38){$\scr{j_l}$}
\pspolygon[](38,45)(40,43)(38,41)(36,43)
\psline[fillstyle=solid](49,32)(51.83,29.17)
\pspolygon[](48,35)(50,33)(48,31)(46,33)
\rput{43.83}(32.44,48.44){\psellipse[](0,0)(2.21,-2.12)}
\psline[fillstyle=solid](11,70)(16.16,64.84)
\psline{|*-}(80,57)(77.17,54.17)
\rput(64,38){$\scr{j_l}$}
\rput{45}(70.5,47.5){\psellipse[](0,0)(2.21,-2.12)}
\rput(75,46){$\scr{h^{-1}_{l_x}}$}
\psline[fillstyle=solid]{-|}(56.17,33.17)(59,36)
\pspolygon[](53,32)(55,34)(57,32)(55,30)
\psline[fillstyle=solid](52,29)(54.16,31.16)
\rput(58,28){$\scr{R_x}$}
\psline[fillstyle=solid]{|*-}(82.17,59.17)(85,62)
\rput(88,58){$\scr{R^{-1}_x}$}
\rput(80,52){$\scr{R_x}$}
\psline[fillstyle=solid](72.17,49.17)(75,52)
\psline(69,46)(66.18,43.18)
\pspolygon[](63,42)(65,44)(67,42)(65,40)
\rput(71,40){$\scr{R^{-1}_x}$}
\psline[fillstyle=solid]{|-}(61,38)(63.83,40.83)
\pspolygon[](74,53)(76,55)(78,53)(76,51)
\psline[fillstyle=solid](87,64)(92,69)
\pspolygon[](84,63)(86,65)(88,63)(86,61)
\rput(41,108){$\scr{j_x}$}
\rput(58,108){$\scr{j_y}$}
\rput(101,67){$\scr{j'_y}$}
\rput(3,68){$\scr{j'_x}$}
\end{pspicture}
\ea
\Bigg\rangle
\label{psi}
\ee

Thanks to \eqref{nuova H} the expectation value of the scalar constraint on such state is given by 
\be
\begin{split}
&\langle\Psi_{H} \; \vgraph^z |{}^{R}\hat{H}^\vgraph_{E\cube} |\Psi_{H} \; \vgraph^z\rangle=
-4i (8\pi\gamma l_P^2)^{\frac{1}{2}}\mathcal{N}(\vgraph)\sum_{j_x,j_y,j_z, j_l}\;\sum_{\mu'_x,\mu'_y,\mu_x,\mu_y=\pm\frac{1}{2}}\sum_{\mu=\pm\frac{1}{2}} \sqrt{j_x\;j_y\;(j_z+\mu)}\; s(\mu)  C^{1\,0}_{\frac{1}{2}\,\frac{1}{2} \; \frac{1}{2}\,-\frac{1}{2}}
\\
&
\ba
\ifx\JPicScale\undefined\def\JPicScale{0.7}\fi
\psset{unit=\JPicScale mm}
\psset{linewidth=0.3,dotsep=1,hatchwidth=0.3,hatchsep=1.5,shadowsize=1,dimen=middle}
\psset{dotsize=0.7 2.5,dotscale=1 1,fillcolor=black}
\psset{arrowsize=1 2,arrowlength=1,arrowinset=0.25,tbarsize=0.7 5,bracketlength=0.15,rbracketlength=0.15}
\begin{pspicture}(0,0)(104,120)
\psline[fillstyle=solid]{-|}(42.83,101.83)(40,99)
\pspolygon[](46,103)(44,101)(42,103)(44,105)
\rput(40,104){$\scr{R_x}$}
\rput(43,108){$\scr{j_x}$}
\rput(13,77){$\scr{j_x}$}
\psline[fillstyle=solid]{-|}(51,110)(51,120)
\psline[fillstyle=solid]{-|}(59.17,101.83)(62,99)
\rput(62,104){$\scr{R_y}$}
\rput(57,107){$\scr{j_y}$}
\rput(90,77){$\scr{j_y}$}
\psline[fillstyle=solid]{|*-}(44.17,95.17)(47,98)
\rput(48,94){$\scr{R_x}$}
\rput(51,104){$\scr{\frac{1}{2}}$}
\psline[fillstyle=solid](49,100)(53,104)
\pspolygon[](46,99)(48,101)(50,99)(48,97)
\psline[fillstyle=solid]{|*-}(57.83,95.17)(55,98)
\rput(55,94){$\scr{R_y^{-1}}$}
\rput(60,111){$\scr{1}$}
\pspolygon[](54,97)(52,99)(54,101)(56,99)
\psline[fillstyle=solid](53,104)(61,115)
\psline[fillstyle=solid](54,101)(53,104)
\rput(49,112){$\scr{j_z}$}
\rput(47,120){$\scr{j_z}$}
\psline[border=0.3,fillstyle=solid](51,110)(57,104)
\psline[fillstyle=solid](51,110)(44.84,103.84)
\pspolygon[](58,105)(60,103)(58,101)(56,103)
\rput(62,116){$\scr{0}$}
\psline[fillstyle=solid]{|*-}(16.83,75.83)(14,73)
\rput(10,74){$\scr{R^{-1}_x}$}
\psline[fillstyle=solid](12,71)(-2,57)
\pspolygon[](15,72)(13,70)(11,72)(13,74)
\psline[fillstyle=solid]{|*-}(85.17,75.83)(88,73)
\rput(95,73){$\scr{R^{-1}_y}$}
\psline[fillstyle=solid](90,71)(104,57)
\pspolygon[](89,74)(91,72)(89,70)(87,72)
\rput(11,46){$\scr{j_l}$}
\rput(93,50){$\scr{j_l}$}
\psline[fillstyle=solid]{-|}(18.17,49.83)(21,47)
\pspolygon[](17,53)(19,51)(17,49)(15,51)
\rput(9,50){$\scr{R_y}$}
\psline[fillstyle=solid]{|*-}(44.17,23.83)(47,21)
\rput(44,17){$\scr{R^{-1}_y}$}
\rput(52,12){$\scr{j_l}$}
\psline[fillstyle=solid](49,19)(51.83,16.17)
\pspolygon[](48,22)(50,20)(48,18)(46,20)
\psline[fillstyle=solid](4,63)(16.16,51.84)
\psline[fillstyle=solid]{-|}(56.17,20.17)(59,23)
\pspolygon[](53,19)(55,21)(57,19)(55,17)
\psline[fillstyle=solid](52,16)(54.16,18.16)
\rput(58,15){$\scr{R_x}$}
\psline[fillstyle=solid]{|*-}(82.17,46.17)(85,49)
\rput(88,45){$\scr{R^{-1}_x}$}
\psline[fillstyle=solid](87,51)(99,62)
\pspolygon[](84,50)(86,52)(88,50)(86,48)
\rput(104,62){$\scr{j'_y}$}
\rput(-2,62){$\scr{j'_x}$}
\psline[fillstyle=solid]{|*-}(18.83,70.83)(16,68)
\rput(21,67){$\scr{R^{-1}_x}$}
\rput(25,72){$\scr{\mu_x}$}
\psline[fillstyle=solid](14,66)(13,65)
\pspolygon[](17,67)(15,65)(13,67)(15,69)
\psline[fillstyle=solid]{|*-}(83.17,70.83)(86,68)
\rput(80,68){$\scr{R_y}$}
\rput(78,72){$\scr{\mu_y}$}
\psline[fillstyle=solid](88,66)(90,64)
\pspolygon[](87,69)(89,67)(87,65)(85,67)
\rput(24,59){$\scr{\mu'_y}$}
\rput(77,57){$\scr{\mu'_x}$}
\rput(43,35){$\scr{\mu'_y}$}
\psline[fillstyle=solid]{-|}(18.17,59.83)(21,57)
\pspolygon[](17,63)(19,61)(17,59)(15,61)
\rput(21,62){$\scr{R_y}$}
\psline[fillstyle=solid]{|*-}(44.17,33.83)(47,31)
\rput(49,34){$\scr{R^{-1}_y}$}
\psline[fillstyle=solid](49,29)(51.83,26.17)
\pspolygon[](48,32)(50,30)(48,28)(46,30)
\psline[fillstyle=solid](13,65)(16.16,61.84)
\rput(56,37){$\scr{\mu'_x}$}
\psline[fillstyle=solid]{-|}(56.17,30.17)(59,33)
\pspolygon[](53,29)(55,31)(57,29)(55,27)
\psline[fillstyle=solid](52,26)(54.16,28.16)
\rput(55,34){$\scr{R_x}$}
\psline[fillstyle=solid]{|*-}(82.17,56.17)(85,59)
\rput(80,62){$\scr{R^{-1}_x}$}
\psline[fillstyle=solid](87,61)(90,64)
\pspolygon[](84,60)(86,62)(88,60)(86,58)
\rput(42,93){$\scr{\mu_x}$}
\rput(60,93){$\scr{\mu_y}$}
\end{pspicture}
\ea
\hspace{1cm}
\ba
\ifx\JPicScale\undefined\def\JPicScale{0.7}\fi
\psset{unit=\JPicScale mm}
\psset{linewidth=0.3,dotsep=1,hatchwidth=0.3,hatchsep=1.5,shadowsize=1,dimen=middle}
\psset{dotsize=0.7 2.5,dotscale=1 1,fillcolor=black}
\psset{arrowsize=1 2,arrowlength=1,arrowinset=0.25,tbarsize=0.7 5,bracketlength=0.15,rbracketlength=0.15}
\begin{pspicture}(0,0)(104,120)
\psline[fillstyle=solid]{-|}(42.83,101.83)(40,99)
\pspolygon[](46,103)(44,101)(42,103)(44,105)
\rput(40,104){$\scr{R_x}$}
\rput(35,101){$\scr{j_x+\mu_x}$}
\rput(13,77){$\scr{j_x+\mu_x}$}
\psline[fillstyle=solid]{-|}(51,110)(51,120)
\psline[fillstyle=solid]{-|}(59.17,101.83)(62,99)
\rput(62,104){$\scr{R_y}$}
\rput(66,102){$\scr{j_y-\mu_y}$}
\rput(90,77){$\scr{j_y-\mu_y}$}
\rput(51,104){$\scr{\frac{1}{2}}$}
\rput(49,112){$\scr{j_z}$}
\rput(47,120){$\scr{j_z}$}
\psline[border=0.3,fillstyle=solid](51,110)(57,104)
\psline[fillstyle=solid](51,110)(44.84,103.84)
\pspolygon[](58,105)(60,103)(58,101)(56,103)
\psline[fillstyle=solid]{|*-}(16.83,75.83)(14,73)
\rput(10,74){$\scr{R^{-1}_x}$}
\psline[fillstyle=solid](12,71)(-2,57)
\pspolygon[](15,72)(13,70)(11,72)(13,74)
\psline[fillstyle=solid]{|*-}(85.17,75.83)(88,73)
\rput(95,73){$\scr{R^{-1}_y}$}
\psline[fillstyle=solid](90,71)(104,57)
\pspolygon[](89,74)(91,72)(89,70)(87,72)
\rput(11,46){$\scr{j_l+\mu'_y}$}
\rput(93,50){$\scr{j_l+\mu'_y}$}
\psline[fillstyle=solid]{-|}(18.17,49.83)(21,47)
\pspolygon[](17,53)(19,51)(17,49)(15,51)
\rput(9,50){$\scr{R_y}$}
\psline[fillstyle=solid]{|*-}(44.17,23.83)(47,21)
\rput(44,17){$\scr{R^{-1}_y}$}
\rput(41,20){$\scr{j_l+\mu'_y}$}
\psline[fillstyle=solid](49,19)(51.83,16.17)
\pspolygon[](48,22)(50,20)(48,18)(46,20)
\psline[fillstyle=solid](4,63)(16.16,51.84)
\psline[fillstyle=solid]{-|}(56.17,20.17)(59,23)
\pspolygon[](53,19)(55,21)(57,19)(55,17)
\psline[fillstyle=solid](52,16)(54.16,18.16)
\rput(58,15){$\scr{R_x}$}
\psline[fillstyle=solid]{|*-}(82.17,46.17)(85,49)
\rput(88,45){$\scr{R^{-1}_x}$}
\psline[fillstyle=solid](87,51)(99,62)
\pspolygon[](84,50)(86,52)(88,50)(86,48)
\rput(104,62){$\scr{j'_y}$}
\rput(-2,62){$\scr{j'_x}$}
\rput(61,19){$\scr{j_l+\mu'_y}$}
\rput(54,113){{\large{*}}}
\rput(84,68){{\large{*}}}
\rput(20,70){{\large{*}}}
\rput(52,25){{\large{*}}}
\end{pspicture}
\ea
\\
&
\left(
\ba
\ifx\JPicScale\undefined\def\JPicScale{0.6}\fi
\psset{unit=\JPicScale mm}
\psset{linewidth=0.3,dotsep=1,hatchwidth=0.3,hatchsep=1.5,shadowsize=1,dimen=middle}
\psset{dotsize=0.7 2.5,dotscale=1 1,fillcolor=black}
\psset{arrowsize=1 2,arrowlength=1,arrowinset=0.25,tbarsize=0.7 5,bracketlength=0.15,rbracketlength=0.15}
\begin{pspicture}(0,0)(101,143)
\rput(28,87){$\Psi_{H_{l_x}}(j_x+\mu_x)$}
\psline[fillstyle=solid]{-|}(42.83,101.83)(40,99)
\pspolygon[](46,103)(44,101)(42,103)(44,105)
\psline[fillstyle=solid](51,110)(44.84,103.84)
\rput(40,104){$\scr{R_x}$}
\psline[fillstyle=solid]{|*-}(16.83,75.83)(14,73)
\rput(10,74){$\scr{R^{-1}_x}$}
\rput(11,77){$\scr{j_x+\mu_x}$}
\psline[fillstyle=solid]{-|}(51,110)(51,120)
\psline[fillstyle=solid]{|*-}(51,139)(51,143)
\rput(46,141){$\scr{j_z}$}
\psline[fillstyle=solid](12,71)(6,65)
\pspolygon[](15,72)(13,70)(11,72)(13,74)
\psline[fillstyle=solid]{-|}(59.17,101.83)(62,99)
\pspolygon[](58,105)(60,103)(58,101)(56,103)
\rput(62,104){$\scr{R_y}$}
\psline[fillstyle=solid]{|*-}(85.17,75.83)(88,73)
\rput(95,73){$\scr{R^{-1}_y}$}
\rput(90,77){$\scr{j_y-\mu_y}$}
\psline[fillstyle=solid](90,71)(96,65)
\pspolygon[](89,74)(91,72)(89,70)(87,72)
\rput(14,60){$\scr{j_l+\mu'_y}$}
\rput(80,64){$\scr{j_l+\mu'_x}$}
\psline[border=0.3,fillstyle=solid](51,110)(57.16,103.84)
\rput(47,120){$\scr{j_z}$}
\psline[fillstyle=solid]{-|}(18.17,62.83)(21,60)
\pspolygon[](17,66)(19,64)(17,62)(15,64)
\rput(21,65){$\scr{R_y}$}
\psline[fillstyle=solid]{|*-}(44.17,36.83)(47,34)
\rput(44,30){$\scr{R^{-1}_y}$}
\rput(45,40){$\scr{j_l+\mu'_y}$}
\rput(63,40){$\scr{j_l+\mu'_x}$}
\psline[fillstyle=solid](49,32)(51.83,29.17)
\pspolygon[](48,35)(50,33)(48,31)(46,33)
\psline[fillstyle=solid](11,70)(16.16,64.84)
\psline[fillstyle=solid]{-|}(56.17,33.17)(59,36)
\pspolygon[](53,32)(55,34)(57,32)(55,30)
\psline[fillstyle=solid](52,29)(54.16,31.16)
\rput(58,28){$\scr{R_x}$}
\psline[fillstyle=solid]{|*-}(82.17,59.17)(85,62)
\psline[fillstyle=solid](87,64)(92,69)
\pspolygon[](84,63)(86,65)(88,63)(86,61)
\rput(41,108){$\scr{j_x+\mu_x}$}
\rput(58,108){$\scr{j_y-\mu_y}$}
\rput(101,67){$\scr{j'_y}$}
\rput(3,68){$\scr{j'_x}$}
\rput(73,87){$\Psi_{H_{l_y}}(j_y-\mu_y)$}
\rput(33,48){$\Psi_{H_{l_{l_{y}}}}(j_l+\mu_y')$}
\rput(71,48){$\Psi_{H_{l_{l_{x}}}}(j_l+\mu_x')$}
\rput(51,130){$\Psi_{H_{l_z}}(j_z)$}
\rput(54,113){{\large{*}}}
\rput(84,68){{\large{*}}}
\rput(20,70){{\large{*}}}
\rput(52,25){{\large{*}}}
\end{pspicture}
\ea
\right)^{{\huge{*}}}
\left(\ba
\ifx\JPicScale\undefined\def\JPicScale{0.6}\fi
\psset{unit=\JPicScale mm}
\psset{linewidth=0.3,dotsep=1,hatchwidth=0.3,hatchsep=1.5,shadowsize=1,dimen=middle}
\psset{dotsize=0.7 2.5,dotscale=1 1,fillcolor=black}
\psset{arrowsize=1 2,arrowlength=1,arrowinset=0.25,tbarsize=0.7 5,bracketlength=0.15,rbracketlength=0.15}
\begin{pspicture}(0,0)(101,143)
\rput(28,87){$\Psi_{H_{l_x}}(j_x)$}
\psline[fillstyle=solid]{-|}(42.83,101.83)(40,99)
\pspolygon[](46,103)(44,101)(42,103)(44,105)
\psline[fillstyle=solid](51,110)(44.84,103.84)
\rput(40,104){$\scr{R_x}$}
\psline[fillstyle=solid]{|*-}(16.83,75.83)(14,73)
\rput(10,74){$\scr{R^{-1}_x}$}
\rput(11,77){$\scr{j_x}$}
\psline[fillstyle=solid]{-|}(51,110)(51,120)
\psline[fillstyle=solid]{|*-}(51,139)(51,143)
\rput(46,141){$\scr{j_z}$}
\psline[fillstyle=solid](12,71)(6,65)
\pspolygon[](15,72)(13,70)(11,72)(13,74)
\psline[fillstyle=solid]{-|}(59.17,101.83)(62,99)
\pspolygon[](58,105)(60,103)(58,101)(56,103)
\rput(62,104){$\scr{R_y}$}
\psline[fillstyle=solid]{|*-}(85.17,75.83)(88,73)
\rput(95,73){$\scr{R^{-1}_y}$}
\rput(90,77){$\scr{j_y}$}
\psline[fillstyle=solid](90,71)(96,65)
\pspolygon[](89,74)(91,72)(89,70)(87,72)
\rput(14,60){$\scr{j_l}$}
\rput(80,64){$\scr{j_l}$}
\psline[border=0.3,fillstyle=solid](51,110)(57.16,103.84)
\rput(47,120){$\scr{j_z}$}
\psline[fillstyle=solid]{-|}(18.17,62.83)(21,60)
\pspolygon[](17,66)(19,64)(17,62)(15,64)
\rput(21,65){$\scr{R_y}$}
\psline[fillstyle=solid]{|*-}(44.17,36.83)(47,34)
\rput(44,30){$\scr{R^{-1}_y}$}
\rput(49,38){$\scr{j_l}$}
\psline[fillstyle=solid](49,32)(51.83,29.17)
\pspolygon[](48,35)(50,33)(48,31)(46,33)
\psline[fillstyle=solid](11,70)(16.16,64.84)
\psline[fillstyle=solid]{-|}(56.17,33.17)(59,36)
\pspolygon[](53,32)(55,34)(57,32)(55,30)
\psline[fillstyle=solid](52,29)(54.16,31.16)
\rput(58,28){$\scr{R_x}$}
\psline[fillstyle=solid]{|*-}(82.17,59.17)(85,62)
\psline[fillstyle=solid](87,64)(92,69)
\pspolygon[](84,63)(86,65)(88,63)(86,61)
\rput(41,108){$\scr{j_x}$}
\rput(58,108){$\scr{j_y}$}
\rput(101,67){$\scr{j'_y}$}
\rput(3,68){$\scr{j'_x}$}
\rput(73,87){$\Psi_{H_{l_y}}(j_y)$}
\rput(33,48){$\Psi_{H_{l_l{y}}}(j_l)$}
\rput(71,48){$\Psi_{H_{l_l{x}}}(j_l)$}
\rput(51,130){$\Psi_{H_{l_z}}(j_z)$}
\rput(54,113){{\large{*}}}
\rput(84,68){{\large{*}}}
\rput(20,70){{\large{*}}}
\rput(52,25){{\large{*}}}
\end{pspicture}
\ea\right)
\end{split}
\label{azione H esatta}
\ee

The expression above can be simplified using the fact that the intertwiners appearing should be all normalized. This can be done by dividing each three valent node by

\be
\sqrt{|<{\bf j_l}, {\bf x_{n_3}}|{\bf n_l}, \vec{{\bf u}}_l >|^2}=\sqrt{
\left(
\begin{array} {c}
\ifx\JPicScale\undefined\def\JPicScale{0.7}\fi
\psset{unit=\JPicScale mm}
\psset{linewidth=0.3,dotsep=1,hatchwidth=0.3,hatchsep=1.5,shadowsize=1,dimen=middle}
\psset{dotsize=0.7 2.5,dotscale=1 1,fillcolor=black}
\psset{arrowsize=1 2,arrowlength=1,arrowinset=0.25,tbarsize=0.7 5,bracketlength=0.15,rbracketlength=0.15}
\begin{pspicture}(7,0)(32,27)
\psline[fillstyle=solid]{-|}(12.83,8.83)(10,6)
\pspolygon[](16,10)(14,8)(12,10)(14,12)
\psline[fillstyle=solid](21,17)(14.84,10.84)
\rput(10,11){$\scr{R_1}$}
\psline[fillstyle=solid]{-|}(21,24)(21,27)
\psline[fillstyle=solid]{-|}(29.17,8.83)(32,6)
\pspolygon[](28,12)(30,10)(28,8)(26,10)
\rput(32,11){$\scr{R_2}$}
\psline[fillstyle=solid](21,17)(27.16,10.84)
\rput(17,27){$\scr{j_3}$}
\rput(11,15){$\scr{j_1}$}
\rput(28,15){$\scr{j_2}$}
\pspolygon[](19.5,24)(22.5,24)(22.5,21)(19.5,21)
\psline(21,21)(21,17)
\rput(25,22){$\scr{R_3}$}
\end{pspicture}
\end{array}
\right)^{*}
\begin{array} {c}
\ifx\JPicScale\undefined\def\JPicScale{0.7}\fi
\psset{unit=\JPicScale mm}
\psset{linewidth=0.3,dotsep=1,hatchwidth=0.3,hatchsep=1.5,shadowsize=1,dimen=middle}
\psset{dotsize=0.7 2.5,dotscale=1 1,fillcolor=black}
\psset{arrowsize=1 2,arrowlength=1,arrowinset=0.25,tbarsize=0.7 5,bracketlength=0.15,rbracketlength=0.15}
\begin{pspicture}(0,0)(32,27)
\psline[fillstyle=solid]{-|}(12.83,8.83)(10,6)
\pspolygon[](16,10)(14,8)(12,10)(14,12)
\psline[fillstyle=solid](21,17)(14.84,10.84)
\rput(10,11){$\scr{R_1}$}
\psline[fillstyle=solid]{-|}(21,24)(21,27)
\psline[fillstyle=solid]{-|}(29.17,8.83)(32,6)
\pspolygon[](28,12)(30,10)(28,8)(26,10)
\rput(32,11){$\scr{R_2}$}
\psline[fillstyle=solid](21,17)(27.16,10.84)
\rput(17,27){$\scr{j_3}$}
\rput(11,15){$\scr{j_1}$}
\rput(28,15){$\scr{j_2}$}
\pspolygon[](19.5,24)(22.5,24)(22.5,21)(19.5,21)
\psline(21,21)(21,17)
\rput(25,22){$\scr{R_3}$}
\end{pspicture}
\end{array}
}
\ee 
In the coherent states \eqref{psi}, this normalization must be done twice: for both the intertwiners in the basis elements and the intertwiners in the coefficients (since the latter are dual to the former). This corresponds to use a normalized intertwiner basis, for which each intertwiner is just a phase and the expression above is equal to 1.  
Having normalized intertwiners, the full state $|\Psi_{H} \; \vgraph \rangle$ is normalized too, {\it i.e.}
\be
\langle \Psi_{H} \; \vgraph\,  |\Psi_{H} \; \vgraph \,\rangle\,=1\,.
\ee
and \eqref{azione H esatta} becomes

\be
\begin{split}
&\langle\Psi_{H} \; \vgraph^z |{}^{R}\hat{H}^\vgraph_{E\cube} |\Psi_{H} \; \vgraph^z\rangle= -4i (8\pi\gamma l_P^2)^{\frac{1}{2}}\mathcal{N}(\vgraph)\\
&\sum_{\mu=\pm \frac{1}{2}}\sum_{\mu_x,\mu_y=\pm \frac{1}{2}}\sum_{\mu'_x,\mu'_y=\pm \frac{1}{2}}\sum_{j_x,j_y,j_z,j_l} \sqrt{j_x\;j_y\;(j_z+\mu)}\; s(\mu)  C^{1\,0}_{\frac{1}{2}\,\frac{1}{2} \; \frac{1}{2}\,-\frac{1}{2}}
\ba
\ifx\JPicScale\undefined\def\JPicScale{0.6}\fi
\psset{unit=\JPicScale mm}
\psset{linewidth=0.3,dotsep=1,hatchwidth=0.3,hatchsep=1.5,shadowsize=1,dimen=middle}
\psset{dotsize=0.7 2.5,dotscale=1 1,fillcolor=black}
\psset{arrowsize=1 2,arrowlength=1,arrowinset=0.25,tbarsize=0.7 5,bracketlength=0.15,rbracketlength=0.15}
\begin{pspicture}(0,0)(101,143)
\rput(18,87){$\Psi^*_{H_{l_x}}(j_x+\mu_x)\Psi_{H_{l_x}}(j_x)$}
\psline[fillstyle=solid]{-|}(42.83,101.83)(40,99)
\pspolygon[](46,103)(44,101)(42,103)(44,105)
\psline[fillstyle=solid](51,110)(44.84,103.84)
\rput(40,104){$\scr{R_x}$}
\psline[fillstyle=solid]{|*-}(16.83,75.83)(14,73)
\rput(10,74){$\scr{R^{-1}_x}$}
\rput(19,78){$\scr{\mu_x}$}
\psline[fillstyle=solid]{-|}(51,110)(51,120)
\psline[fillstyle=solid](12,71)(11,70)
\pspolygon[](15,72)(13,70)(11,72)(13,74)
\psline[fillstyle=solid]{-|}(59.17,101.83)(62,99)
\pspolygon[](58,105)(60,103)(58,101)(56,103)
\rput(62,104){$\scr{R_y}$}
\psline[fillstyle=solid]{|*-}(85.17,75.83)(88,73)
\rput(95,73){$\scr{R^{-1}_y}$}
\rput(84,78){$\scr{\mu_y}$}
\psline[fillstyle=solid](90,71)(92,69)
\pspolygon[](89,74)(91,72)(89,70)(87,72)
\rput(11,64){$\scr{\frac{1}{2}}$}
\rput(22,57){$\scr{\mu'_y}$}
\rput(93,66){$\scr{\frac{1}{2}}$}
\rput(81,56){$\scr{\mu'_x}$}
\psline[border=0.3,fillstyle=solid](51,110)(57.16,103.84)
\rput(51,122){$\scr{0}$}
\psline[fillstyle=solid]{-|}(18.17,62.83)(21,60)
\pspolygon[](17,66)(19,64)(17,62)(15,64)
\rput(21,65){$\scr{R_y}$}
\psline[fillstyle=solid]{|*-}(44.17,36.83)(47,34)
\rput(44,30){$\scr{R^{-1}_y}$}
\rput(53,27){$\scr{\frac{1}{2}}$}
\rput(40,38){$\scr{\mu'_y}$}
\rput(61,38){$\scr{\mu'_x}$}
\psline[fillstyle=solid](49,32)(51.83,29.17)
\pspolygon[](48,35)(50,33)(48,31)(46,33)
\psline[fillstyle=solid](11,70)(16.16,64.84)
\psline[fillstyle=solid]{-|}(56.17,33.17)(59,36)
\pspolygon[](53,32)(55,34)(57,32)(55,30)
\psline[fillstyle=solid](52,29)(54.16,31.16)
\rput(58,28){$\scr{R_x}$}
\rput(92,60){$\scr{R_x}^{-1}$}
\psline[fillstyle=solid]{|*-}(82.17,59.17)(85,62)
\psline[fillstyle=solid](87,64)(92,69)
\pspolygon[](84,63)(86,65)(88,63)(86,61)
\rput(48,115){$\scr{1}$}
\rput(45,108){$\scr{\frac{1}{2}}$}
\rput(38,97){$\scr{\mu_x}$}
\rput(57,108){$\scr{\frac{1}{2}}$}
\rput(65,97){$\scr{\mu_y}$}
\rput(88,87){$\Psi^*_{H_{l_y}}(j_y-\mu_y)\Psi_{H_{l_y}}(j_y)$}
\rput(18,48){$\Psi^*_{H_{l_{l_{y}}}}(j_l+\mu'_y)\Psi_{H_{l_{l_{y}}}}(j_l)$}
\rput(88,48){$\Psi^*_{H_{l_{l_{x}}}}(j_l+\mu'_x)\Psi_{H_{l_{l_{x}}}}(j_l)$}
\rput(51,130){$\Psi^*_{H_{l_z}}(j_z)\Psi_{H_{l_z}}(j_z)$}
\end{pspicture}
\ea
\end{split}
\label{LO1}
\ee
This expression coincides with the leading order term of Eq.(71) in \cite{Alesci:2014uha} modulo the coefficient, but here no large $j$'s limit has been performed yet. 

Hence, we can use our previous results to evaluate \eqref{LO1}. In particular, taking the variances $\alpha_l=1/(\bar{j}_l)^k$ with $k>1$ \cite{Bianchi:2009ky}, we can perform a saddle point expansion around the expectation values $\bar{j}_l$ of the functions $\Psi_H$ (which are almost gaussian), while the corrections are order $\bar{j}^{-k}_l$, thus negligible in the large $j$'s limit. 

This way, one gets
\be
\langle\Psi_{H} \; \vgraph^z|{}^{R}\hat{H}^\vgraph_{E\cube} |\Psi_{H} \; \vgraph^z\rangle\approx
2\mathcal{N}(\vgraph)(8\pi\gamma l_P^2)^{1/2}\sum_{\mu=\pm 1/2}\sqrt{\bar{j}_x\;\bar{j}_y\;(\bar{j}_z+\mu)}\; s(\mu) \sin{\bar{c}_x} \sin{\bar{c}_y},
\ee 
and the only differences with Eq.(74) in \cite{Alesci:2014uha} are that we are now using physical phase variables and we have a different coefficient. We can expand the square root $\sqrt{\bar{j}_z+\mu}$ for $\bar{j}_z\gg 1/2$, so getting
\be
\sum_{\mu=\pm 1/2}\sqrt{\bar{j}_z+\mu}\; s(\mu)\approx \frac{1}{\bar{j}_z} \left[1+\frac{1}{(8\bar{j}_z)^2}\right]\,,\label{ivexp}
\ee
and by inserting physical momenta $\bar{p}_i$ for a single cell (smeared on the dual surfaces to the links $l_i$) one gets
\be
\langle\Psi_{H} \; \vgraph^z|{}^{R}\hat{H}^\vgraph_{E\cube} |\Psi_{H} \; \vgraph^z\rangle\approx
2\mathcal{N}(\vgraph)\sqrt{\frac{\bar{p}_x\;\bar{p}_y}{\bar{p}_z}}\;  \sin{\bar{c}_x} \sin{\bar{c}_y}\,.\label{1ham}
\ee 

It is worth noting how such a new large $j$'s expansion provides the so-called inverse-volume corrections, which are of $1/j$ order, thus generically bigger than those related with the saddle point approximation for gaussians. 

The extension to the six-valent case is straightforward. The action of the scalar constraint is the same as in the three-valent case. The only difference is that now one has to use  
the normalization of six-valent reduced intertwiners:

\be
\sqrt{|<{\bf j_l}, {\bf x_{n_6}}|{\bf n_l}, \vec{{\bf u}}_l >|^2}=
\sqrt{
\left(
\ba
\ifx\JPicScale\undefined\def\JPicScale{1}\fi
\psset{unit=\JPicScale mm}
\psset{linewidth=0.3,dotsep=1,hatchwidth=0.3,hatchsep=1.5,shadowsize=1,dimen=middle}
\psset{dotsize=0.7 2.5,dotscale=1 1,fillcolor=black}
\psset{arrowsize=1 2,arrowlength=1,arrowinset=0.25,tbarsize=0.7 5,bracketlength=0.15,rbracketlength=0.15}
\begin{pspicture}(50,50)(87,70)
\psline[fillstyle=solid]{-|}(62.83,49.83)(60,47)
\pspolygon[](66,51)(64,49)(62,51)(64,53)
\rput(66,48){$\scr{R_x}$}
\psline[fillstyle=solid]{-|}(71,58)(71,68)
\psline[fillstyle=solid]{-|}(83,58)(87,58)
\rput(83,62){$\scr{R_y}$}
\rput(84,55){$\scr{j_{\vec{n},y}}$}
\rput(67,68){$\scr{j_{\vec{n},z}}$}
\psline[fillstyle=solid](71,58)(64.84,51.84)
\pspolygon[](79.93,59.41)(82.76,59.41)(82.76,56.59)(79.93,56.59)
\psline[fillstyle=solid](71.51,58)(80,58)
\psline{-|}(71,58)(71,47)
\rput(64,62){$\scr{R^{-1}_y}$}
\rput(57,56){$\scr{j_{\vec{n}-\vec{e_y},y}}$}
\psline[fillstyle=solid]{-|}(60,58)(56,58)
\pspolygon[](63.07,56.59)(60.24,56.59)(60.24,59.41)(63.07,59.41)
\psline[fillstyle=solid](71.49,58)(63,58)
\psline[fillstyle=solid]{-|}(79.17,66.17)(82,69)
\pspolygon[](76,65)(78,67)(80,65)(78,63)
\rput(78,70){$\scr{R^{-1}_x}$}
\rput(86,66){$\scr{j_{\vec{n}-\vec{e_x},x}}$}
\psline[fillstyle=solid](71,58)(77.16,64.16)
\rput(78,47){$\scr{j_{\vec{n}-\vec{e}_z,z}}$}
\rput(58,51){$\scr{j_{\vec{n},x}}$}
\end{pspicture}
\ea\quad
\right)^{*}
\ba
\ifx\JPicScale\undefined\def\JPicScale{1}\fi
\psset{unit=\JPicScale mm}
\psset{linewidth=0.3,dotsep=1,hatchwidth=0.3,hatchsep=1.5,shadowsize=1,dimen=middle}
\psset{dotsize=0.7 2.5,dotscale=1 1,fillcolor=black}
\psset{arrowsize=1 2,arrowlength=1,arrowinset=0.25,tbarsize=0.7 5,bracketlength=0.15,rbracketlength=0.15}
\begin{pspicture}(50,50)(87,70)
\psline[fillstyle=solid]{-|}(62.83,49.83)(60,47)
\pspolygon[](66,51)(64,49)(62,51)(64,53)
\rput(66,48){$\scr{R_x}$}
\psline[fillstyle=solid]{-|}(71,58)(71,68)
\psline[fillstyle=solid]{-|}(83,58)(87,58)
\rput(83,62){$\scr{R_y}$}
\rput(84,55){$\scr{j_{\vec{n},y}}$}
\rput(67,68){$\scr{j_{\vec{n},z}}$}
\psline[fillstyle=solid](71,58)(64.84,51.84)
\pspolygon[](79.93,59.41)(82.76,59.41)(82.76,56.59)(79.93,56.59)
\psline[fillstyle=solid](71.51,58)(80,58)
\psline{-|}(71,58)(71,47)
\rput(64,62){$\scr{R^{-1}_y}$}
\rput(57,56){$\scr{j_{\vec{n}-\vec{e_y},y}}$}
\psline[fillstyle=solid]{-|}(60,58)(56,58)
\pspolygon[](63.07,56.59)(60.24,56.59)(60.24,59.41)(63.07,59.41)
\psline[fillstyle=solid](71.49,58)(63,58)
\psline[fillstyle=solid]{-|}(79.17,66.17)(82,69)
\pspolygon[](76,65)(78,67)(80,65)(78,63)
\rput(78,70){$\scr{R^{-1}_x}$}
\rput(86,66){$\scr{j_{\vec{n}-\vec{e_x},x}}$}
\psline[fillstyle=solid](71,58)(77.16,64.16)
\rput(78,47){$\scr{j_{\vec{n}-\vec{e}_z,z}}$}
\rput(58,51){$\scr{j_{\vec{n},x}}$}
\end{pspicture}
\ea
}
\ee

obtaining
\be
\begin{split}
&\langle\Psi_{H} \; \vgraph_{0,0,0} |{}^{R}\hat{H}^\vgraph_{E\cube} |\Psi_{H} \; \vgraph_{0,0,0}\rangle= -4i(8\pi\gamma l_P^2)^{1/2}\mathcal{N}(\vgraph) 
\sum_{\mu=\pm\frac{1}{2}}\sum_{\mu_x,\mu_y=\pm\frac{1}{2}}\sum_{\mu'_x,\mu'_y=\pm\frac{1}{2}}\sum_{j_x,j_y,j_z,j_{\vec{e}_x,y},j_{\vec{e}_y,x}} \\
& \sqrt{j_x\;j_y\;(j_z+\mu)}\; s(\mu)  C^{1\,0}_{\frac{1}{2}\,\frac{1}{2} \; \frac{1}{2}\,-\frac{1}{2}}
\ba
\ifx\JPicScale\undefined\def\JPicScale{0.6}\fi
\psset{unit=\JPicScale mm}
\psset{linewidth=0.3,dotsep=1,hatchwidth=0.3,hatchsep=1.5,shadowsize=1,dimen=middle}
\psset{dotsize=0.7 2.5,dotscale=1 1,fillcolor=black}
\psset{arrowsize=1 2,arrowlength=1,arrowinset=0.25,tbarsize=0.7 5,bracketlength=0.15,rbracketlength=0.15}
\begin{pspicture}(0,0)(101,143)
\rput(18,87){$\Psi^*_{H_{l_x}}(j_x+\mu_x)\Psi_{H_{l_x}}(j_x)$}
\psline[fillstyle=solid]{-|}(42.83,101.83)(40,99)
\pspolygon[](46,103)(44,101)(42,103)(44,105)
\psline[fillstyle=solid](51,110)(44.84,103.84)
\rput(40,104){$\scr{R_x}$}
\psline[fillstyle=solid]{|*-}(16.83,75.83)(14,73)
\rput(10,74){$\scr{R^{-1}_x}$}
\rput(19,78){$\scr{\mu_x}$}
\psline[fillstyle=solid]{-|}(51,110)(51,120)
\psline[fillstyle=solid](12,71)(11,70)
\pspolygon[](15,72)(13,70)(11,72)(13,74)
\psline[fillstyle=solid]{-|}(59.17,101.83)(62,99)
\pspolygon[](58,105)(60,103)(58,101)(56,103)
\rput(62,104){$\scr{R_y}$}
\psline[fillstyle=solid]{|*-}(85.17,75.83)(88,73)
\rput(95,73){$\scr{R^{-1}_y}$}
\rput(84,78){$\scr{\mu_y}$}
\psline[fillstyle=solid](90,71)(92,69)
\pspolygon[](89,74)(91,72)(89,70)(87,72)
\rput(11,64){$\scr{\frac{1}{2}}$}
\rput(22,57){$\scr{\mu'_y}$}
\rput(93,66){$\scr{\frac{1}{2}}$}
\rput(81,56){$\scr{\mu'_x}$}
\psline[border=0.3,fillstyle=solid](51,110)(57.16,103.84)
\rput(51,122){$\scr{0}$}
\psline[fillstyle=solid]{-|}(18.17,62.83)(21,60)
\pspolygon[](17,66)(19,64)(17,62)(15,64)
\rput(21,65){$\scr{R_y}$}
\psline[fillstyle=solid]{|*-}(44.17,36.83)(47,34)
\rput(44,30){$\scr{R^{-1}_y}$}
\rput(53,27){$\scr{\frac{1}{2}}$}
\rput(40,38){$\scr{\mu'_y}$}
\rput(61,38){$\scr{\mu'_x}$}
\psline[fillstyle=solid](49,32)(51.83,29.17)
\pspolygon[](48,35)(50,33)(48,31)(46,33)
\psline[fillstyle=solid](11,70)(16.16,64.84)
\psline[fillstyle=solid]{-|}(56.17,33.17)(59,36)
\pspolygon[](53,32)(55,34)(57,32)(55,30)
\psline[fillstyle=solid](52,29)(54.16,31.16)
\rput(58,28){$\scr{R_x}$}
\rput(92,60){$\scr{R_x}^{-1}$}
\psline[fillstyle=solid]{|*-}(82.17,59.17)(85,62)
\psline[fillstyle=solid](87,64)(92,69)
\pspolygon[](84,63)(86,65)(88,63)(86,61)
\rput(48,115){$\scr{1}$}
\rput(45,108){$\scr{\frac{1}{2}}$}
\rput(38,97){$\scr{\mu_x}$}
\rput(57,108){$\scr{\frac{1}{2}}$}
\rput(65,97){$\scr{\mu_y}$}
\rput(88,87){$\Psi^*_{H_{l_y}}(j_y-\mu_y)\Psi_{H_{l_y}}(j_y)$}
\rput(10,48){$\Psi^*_{H_{l_{\vec{e}_x,y}}}(j_{\vec{e}_x,y}+\mu'_y)\Psi_{H_{l_{\vec{e}_x,y}}}(j_{\vec{e}_x,y})$}
\rput(98,48){$\Psi^*_{H_{l_{{\vec{e}_y,x}}}}(j_{\vec{e}_y,x}-\mu'_x)\Psi_{H_{l_{{\vec{e}_y,x}}}}(j_{\vec{e}_y,x})$}
\rput(51,130){$\Psi^*_{H_{l_z}}(j_z)\Psi_{H_{l_z}}(j_z)$}
\end{pspicture}
\ea
\end{split}
\label{LOfinal}
\ee
This expression coincides with \eqref{LO1} and the final result of the single-node Hamiltonian is given by \eqref{1ham} summed over ordered permutations of links within a given triple and over all triples around $\vgraph$. On a local homogeneous configuration ({\it i.e.} when the expectation values of spins along a given direction does not change across the node) the first sum just provides twice the sum over ordered permutations $x\rightarrow y \rightarrow z$, while the second simply cancel the factor $c(n)$ in \eqref{Hridotto}. Finally, the expectation value of the single-node Hamiltonian is given by 
\be
\langle\Psi_{H} \; \vgraph_{0,0,0}|\sum_{\cube}{}^{R}\hat{H}^\vgraph_{E\cube} |\Psi_{H} \; \vgraph_{0,0,0}\rangle\approx
4 \mathcal{N}(\vgraph)\bigg(\sqrt{\frac{\bar{p}_x\;\bar{p}_y}{\bar{p}_z}}\;  \sin{\bar{c}_x} \sin{\bar{c}_y} + \sqrt{\frac{\bar{p}_x\;\bar{p}_y}{\bar{p}_z}}\;  \sin{\bar{c}_x} \sin{\bar{c}_y} + \sqrt{\frac{\bar{p}_x\;\bar{p}_y}{\bar{p}_z}}\;  \sin{\bar{c}_x} \sin{\bar{c}_y}\bigg)\,.\label{ham6}
\ee

\section{Collective variables}\label{VI}

Let us consider each Bianchi I local patch filled with $N=N_xN_yN_z$ six-valent nodes, $N_i$ being the number of nodes on a dual fiducial plane. We want to define the phase-space variables $\{C_i,P_j\}$ describing each local patch as a whole. The semiclassical state is constructed by imposing homogeneity for semiclassical expectation values, such $\{\bar{c}_i,\bar{p}_j\}$ are the same for all nodes and the collective (semiclassical) variables are obtained via the following rescaling
\begin{align} 
&P_x=N_y N_z p_x\qquad P_y=N_z N_x p_y\qquad P_z=N_x N_y p_z\,,\label{resc}\\
&C_x=N_x c_x\qquad C_y= N_y c_y\qquad C_z=N_z c_z\,. 
\end{align}
It is worth noting how such a rescaling is due to the fact that we are using physical variables, whose definitions contain the length or the area of the fiducial cell.    

The expectation value of the scalar constraint is the sum over all nodes of the single-node one \eqref{ham6} and, in terms of collective variables, it reads
\begin{align}
{}_N\langle & {}^{R}\hat{H} \rangle_N
\approx\nonumber\\
&\frac{2}{8\pi G\gamma^2}\mathcal{N}\bigg(N_x\, N_y\,\sqrt{\frac{\bar{P}_x\;\bar{P}_y}{\bar{P}_z}}\;  \sin{\frac{\bar{C}_x}{N_x}} \sin{\frac{\bar{C}_y}{N_y}}
+N_y\, N_z\,\sqrt{\frac{\bar{P}_y\;\bar{P}_z}{\bar{P}_x}}\;  \sin{\frac{\bar{C}_y}{N_y}} \sin{\frac{\bar{C}_z}{N_z}}
+N_z\, N_x\,\sqrt{\frac{\bar{P}_z\;\bar{P}_x}{\bar{P}_y}}\;  \sin{\frac{\bar{C}_z}{N_z}} \sin{\frac{\bar{C}_x}{N_x}}\bigg)\,.\label{Nham}
\end{align}
This expression formally coincide with the quantum operator adopted in LQC for the Bianchi I model \cite{MartinBenito:2008wx,Ashtekar:2009vc}, in which one takes as regulators the inverse of the number of nodes on fiducial planes, {\it i.e.} 
\be
\mu_i=\frac{1}{N_i}\,.
\ee 
Therefore, we infer a correspondence between the quantum dynamics of LQC and the semiclassical description of QRLG, similarly to the findings of Group Field Theory cosmology \cite{Gielen:2014uga,Calcagni:2014tga}.  

The regulator is due to the presence of several nodes inside each local patch. The choice of the regularization schemes in LQC correspond to the choice of the regulator as a function of  the momenta variables. This means that the regularization scheme is fixed once we fix the dependence of the number of nodes on $\bar{P}_i$. 

At the same time, the number of nodes together with spin quantum numbers enters the definition of the collective momenta variables $\bar{P}_i$ as follows
\be
\bar{P}_i=\frac{N}{N_i}\, 8\pi\gamma l_P^2\, \bar{j}_i\,.
\ee  
Henceforth, the ambiguity in the regularization is due to the fact that we cannot uniquely fix $N_i$ as a function of $\bar{P}_i$, because another fundamental quantity ($\bar{j}_i$) is involved. For instance, the old regularization scheme is reproduced for a constant number of nodes, such that $\bar{P}_i$ is proportional to $\bar{j}_i$, while the improved scheme can be inferred (with a bigger $\Delta$) for constant spin numbers \cite{Alesci:2014rra}. 
  
The study of corrections reveals that we get different predictions with respect to LQC; while holonomy corrections stay the same, inverse-volume ones differ. This can be seen from the expansion \eqref{ivexp}, which provides the following corrections to the term $1/\sqrt{\bar{p}_z}$ in \eqref{1ham}
\be
\frac{1}{\sqrt{\bar{p}_i}}\rightarrow \frac{1}{\sqrt{\bar{p}_i}}\left[1+\left(\frac{\pi\gamma l_P^2}{\bar{p}_i}\right)^2\right]\,,
\ee  
which, in turn, corresponds to the following modification in \eqref{Nham}
\be
\frac{1}{\sqrt{\bar{P}_i}}\rightarrow \frac{1}{\sqrt{\bar{P}_i}}\left[1+\frac{N^2}{N^2_i}\,\left(\frac{\pi\gamma l_P^2}{\bar{P}_i}\right)^2\right]\,.
\ee
The factor $N$ multiplying the next-to-leading order term implies that we get an enhancement of inverse-volume corrections with respect to LQC. As consequence, the set of admissible regulators is restricted with respect to that compatible with the lattice refinement scheme of LQC \cite{Calcagni} (see \cite{Alesci:2014rra} for details).

\section{Conclusions}\label{VII}

We analyzed the action of the euclidean scalar constraint operator on six-valent nodes in QRLG. This allowed us to investigate a realistic model for the quantum universe described in terms of a discrete graph, whose building blocks are cuboidal cells. We changed the recoupling rule with respect to previous works \cite{Alesci:2013xd,Alesci:2014uha} and we constructed semiclassical states, such that the expectation value of the Hamiltonian equals the classical Hamiltonian in the large $j$ limit. This is a consistency check on the viability of our model and we recognize the new recoupling rule \eqref{newrec} as the correct one. 

We also derived the effective Hamiltonian of LQC \eqref{Nham}, in which the regulator is given by the total number of nodes along fiducial directions within each homogeneous patch. The same result has been obtained in the framework of lattice refinement \cite{Bojowald:2006qu} and in GFT cosmology \cite{Gielen:2014uga,Calcagni:2014tga}. We briefly discussed how different regularization prescriptions can be inferred in our model by assuming that the number of nodes is a function of semiclassical variables. However, the actual implementation of this idea would imply a new definition of the scalar constraint operator, which is behind the scope of the present work. Our analysis is restricted to the non graph-changing version of Thiemann's Hamiltonian, which can be seen as the most conservative definition for the quantum dynamics. Other proposal can be considered, like for instance the one in \cite{Alesci:2010gb}. In this respect, QRLG is a privileged arena to test the phenomenological implications of different definitions of the scalar constraint on a quantum level.   

A distinctive feature of our model with respect to LQC is the magnitude of inverse-volume corrections. These corrections are enhanced, thus they could have a stronger impact on the cosmological observables. In particular, they could significantly affect the dynamics of perturbations and potentially induce sensible modifications to the cosmic microwave background radiation spectrum. For this reason, the study of perturbations is the most urgent issue of QRLG. However, it requires a great theoretical effort, since the dynamics of perturbations is described by those term in the Hamiltonian which are neglected in the homogeneous (or BKL) limit. A first step in this direction is to consider a non-vanishing scalar curvature, adopting the quantization procedure described in \cite{Alesci:2014aza,Alesci:2015wla}.

Future developments will also deal with the introduction of matter fields in the quantum universe described by a collection of six-valent nodes. In this respect, the tools of loop quantization are to be adapted to QRLG framework. If successful, this analysis will provide us with both the effective dynamics of field on a fixed quantum background and the effective contribution to the right-hand side of Einstein equation (within the adopted approximation scheme).

{\acknowledgments
The authors wish to thank G. Calcagni for useful discussions.
The work of FC was supported by funds provided by the National Science Center under the agreement DEC12
2011/02/A/ST2/00294.
The work of E.A. was supported by the grant of Polish Narodowe Centrum Nauki nr 2011/02/A/ST2/00300.}

\end{document}